\documentclass[
11pt,
superscriptaddress,
amsmath,
amssymb,
floatfix,
footinbib,
longbibliography,
notitlepage,
aps,pra,
]{revtex4-1}

\usepackage{graphicx}% Include figure files
\usepackage{dcolumn}% Align table columns on decimal point
\usepackage{bm}% bold math
\usepackage{dsfont}
\usepackage[breaklinks=true,colorlinks,citecolor=blue,linkcolor=blue,urlcolor=blue]{hyperref}% add hypertext capabilities
\usepackage{cleveref}
\usepackage{physics}
\usepackage{mleftright}
\usepackage{bbm}
\usepackage{soul}
\usepackage[euler]{textgreek}

\newcommand{\figref}[1]{\mbox{Fig.~\ref{#1}}}

\newcommand{\secref}[1]{\mbox{Sec.~\ref{#1}}}

\renewcommand{\eqref}[1]{\mbox{Eq.~(\ref{#1})}}

\newcommand{\figpanel}[2]{Fig.~\hyperref[#1]{\ref*{#1}(#2)}}

\newcommand{\be}{\begin{equation}}
\newcommand{\ee}{\end{equation}}
\newcommand{\bea}{\begin{eqnarray}}
\newcommand{\eea}{\end{eqnarray}}

\renewcommand{\eqref}[1]{\mbox{Eq.~(\ref{#1})}}

\usepackage{changes}
\definechangesauthor[name={Ale}, color=red]{Ale}
\definechangesauthor[name={David}, color=orange]{David}
\usepackage{xr}
\usepackage{tikz,xcolor}
\definecolor{lime}{HTML}{A6CE39}
\DeclareRobustCommand{\orcidicon}{%
	\begin{tikzpicture}
	\draw[lime, fill=lime] (0,0) 
	circle [radius=0.16] 
	node[white] {{\fontfamily{qag}\selectfont \tiny ID}};
	\draw[white, fill=white] (-0.0625,0.095) 
	circle [radius=0.007];
	\end{tikzpicture}
	\hspace{-2mm}
}
\foreach \x in {A, ..., Z}{%
	\expandafter\xdef\csname orcid\x\endcsname{\noexpand\href{https://orcid.org/\csname orcidauthor\x\endcsname}{\noexpand\orcidicon}}
}

\begin{document}

%\title{Higher-Order Optomechanical Interactions Account for the Observed Energy Exchange in Optomechanics}
\title{Explanation of the Observed Energy Exchange through the vacuum in Optomechanics}

\author{Vincenzo Macr\`{i}\orcidE{}}
\email{macrivince1978@gmail.com}
\affiliation{Dipartimento di Fisica ``A. Volta", Università di Pavia, Via Bassi 6, 27100 Pavia, Italy}
\author{Franco Nori\orcidF{}}
\affiliation{Center for Quantum Computing, RIKEN, Wakoshi, Saitama, 351-0198, Japan}
\affiliation{Physics Department, The University of Michigan, Ann Arbor, Michigan, 48109-1040, USA}
%\author{Yoshihiko Hasegawa\orcidD{}}
%\affiliation{Department of Information and Communication Engineering, Graduate School of Information Science and Technology, The University of Tokyo}
%\author{Frank K. Wilhelm\orcidC{}}
%\affiliation{Institute for Quantum Computing Analytics (PGI-12), Forschungszentrum J\"ulich, 52425 J\"ulich, Germany}
%\author{David Edward Bruschi\orcidB{}}
%\affiliation{Institute for Quantum Computing Analytics (PGI-12), Forschungszentrum J\"ulich, 52425 J\"ulich, Germany}
\author{Alessandro Ferreri\orcidA{}}
\affiliation{Center for Quantum Computing, RIKEN, Wakoshi, Saitama, 351-0198, Japan}
%\affiliation{Theoretical Quantum Physics Laboratory, RIKEN, Wako-shi, Saitama 351-0198, Japan}

\date{\today}% It is always \today, today,

\begin{abstract}
In cavity optomechanics, the standard large-detuning regime is commonly described by retaining only the radiation-pressure interaction, while higher-order mirror--field interactions are assumed to be negligible. This approximation, however, fails to provide a microscopic explanation for the vacuum-mediated heat transfer observed between the  mechanical membranes in the experiment of Fong \textit{et al.} [\href{https://www.nature.com/articles/s41586-019-1800-4}{Nature.~\textbf{576}, 243 (2019)}], whose physical origin has remained under active debate. Here, using a fully quantum model, we show that the neglected higher-order optomechanical interactions naturally generate phonon-phonon coupling and quantitatively account for the observed energy exchange within the standard optomechanical framework. Building on this microscopic description, we further propose a protocol in which phonon-phonon interaction drives a cyclic process enabling net work extraction. Our results establish the fundamental role of higher-order optomechanical interactions and provide a microscopic framework for describing next-generation optomechanical experiments beyond the linear approximation.
\end{abstract}

\maketitle
%\section*{Introduction}
%\tableofcontents 
{\textbf{Introduction}.---} Over the past two decades, cavity optomechanics has become a leading platform for controlling radiation-pressure interactions between light and matter, enabling quantum control of macroscopic mechanical systems and applications ranging from precision sensing and quantum metrology to nonequilibrium quantum thermodynamics (see, e.g.,~\cite{girvin2florian2009,Aspelmeyer2014,liu2021,barzanjeh2022,Degen2017,Connell2010quantum,Romero2011,brunelli2015,mari_quantum_2015,zhang_quantum_2014,Holmes2020,Tomadin2012,Bibak2023,bragadin2026}). Landmark achievements, including ground-state cooling, quantum backaction, optomechanically induced transparency, and single-phonon quantum control~\cite{groblacher2009,schliesser2009,teufel2011,kippenberg2008,Verlot2010,weis2010,safavi2011,Kronwald2013,Nunnenkamp2011,pirkkalainen2013,rimberg2014,Heikkil2014}, together with recent progress towards the ultrastrong-coupling regime~\cite{frisk_kockum_ultrastrong_2019,Forn2019,Macri2016}, have established cavity optomechanics as a cornerstone of modern quantum science. Most of these advances are described within the standard optomechanical regime, in which optical frequencies greatly exceed mechanical ones~\cite{liu2021,barzanjeh2022}. In this limit, the dynamics is commonly captured by the radiation-pressure interaction, whereas higher-order photon-phonon and phonon-phonon scattering processes arising from the canonical quantization of the coupled field and mirror motion are generally neglected~\cite{law_interaction_1995,Ferreri2022a}. This approximation is well justified for conventional optomechanical configurations, where the large separation between optical and mechanical frequencies would strongly suppress the dynamic effects of such processes.

Recent experimental advances, however, have made both single-photon optomechanical coupling and high-frequency mechanical resonators increasingly accessible~\cite{Connell2010quantum,Ding2010,jiang2012,gil2015,renninger2018,Diamandi2025}. These developments have motivated extensions of the standard framework that incorporates higher-order scattering processes and quadratic mechanical interactions~\cite{Butera22,Butera2025}. Such interactions give rise to a broad range of genuinely quantum phenomena, including the dynamical Casimir effect~\cite{Johansson2009,Wilson2011,Nation2012}, coherent cavity-mediated phonon interactions~\cite{Macri2018,Settineri2019,Butera2019,Ferreri2022a,Ferreri2024,Wang2023,Montalbano2023,Mercurio2025}, and quantum thermodynamic cycles~\cite{Ferreri2023}. 
%These results indicate that higher-order optomechanical processes can generate qualitatively new dynamics beyond the conventional radiation-pressure description.

A particularly relevant configuration is an optical cavity with two movable mirrors, whose full quantum dynamics was investigated numerically in Ref.~\cite{DiStefano2019}. That study predicted phonon exchange mediated by cavity vacuum fluctuations when the mechanical and optical frequencies are comparable, but did not provide a microscopic description of the resulting effective phonon-phonon interaction or address the standard large-detuning regime. The experiments of Fong \textit{et al.}~\cite{fong_phonon_2019}, by contrast, reported vacuum-mediated phonon heat transfer between spatially separated membranes in a regime where the optical frequency greatly exceeds the mechanical one. This observation therefore raises a fundamental question: what microscopic quantum mechanism can account for vacuum-mediated phonon transfer under standard optomechanical conditions?

Here we develop a microscopic theory showing that higher-order optomechanical interactions generate coherent phonon-phonon coupling and correlations, even in the standard large-detuning regime. Our theory provides the first microscopic explanation of the vacuum-mediated heat transfer observed in~\cite{fong_phonon_2019}, identifying the higher-order optomechanical interactions responsible for the observed energy exchange, whose underlying physical mechanism has remained open to different interpretations \cite{biehs2020fundamental}. Building on this framework, we further propose a cyclic protocol in which the same vacuum-mediated phonon conversion enables net work extraction with finite thermodynamic efficiency. Our results establish higher-order optomechanical interactions as an essential ingredient of conventional cavity optomechanics and provide a unified framework for describing vacuum-mediated energy transport and quantum thermodynamic functionalities.

{\textbf{The Unresolved Puzzle of Vacuum-Mediated Heat Transfer}.---} In $2019$ Fong \textit{et al.}~\cite{fong_phonon_2019} experimentally observed a resonant enhancement of thermal energy exchange between two parallel nanomechanical membranes separated by a tunable vacuum gap ($d\sim 200\text{--}1200\,\mathrm{nm}$). The different membrane dimensions allowed their fundamental vibrational modes to be tuned into resonance by thermally adjusting the membrane stress. At bath temperatures $T_1^{\rm bath}=287.0\,\mathrm{K}$ and $T_2^{\rm bath}=312.5\,\mathrm{K}$, resonance was achieved at $\Omega=\Omega_1/2\pi=\Omega_2/2\pi=191.6\,\mathrm{kHz}$, where $\Omega_{(1,2)}$ denote the mechanical frequencies of the two membranes. A striking feature of the experiment is the dependence of the phonon-mode temperatures on the membrane separation. At large distances, the effective phonon temperatures $T'_{(1,2)}$ coincide with the corresponding bath temperatures $T_{(1,2)}^{\rm bath}$. As the separation is reduced below approximately $600\,\mathrm{nm}$, however, the two phonon temperatures progressively deviate from their bath values and get closer to each other. For $d\lesssim400\,\mathrm{nm}$ $T'_1$ and $T'_2$ become almost identical (Fig.~3(a) in~\cite{fong_phonon_2019}), a signature of the thermalization of the two mechanical modes. This behavior was interpreted by the authors as evidence for heat transfer mediated by Casimir vacuum fluctuations.

To support this interpretation, the authors adopted a classical description of the membrane dynamics in which the Casimir force per unit area depends on the distance between the two membranes. This treatment yields an effective Casimir-mediated coupling scaling as  $g_{_\text{\cite{fong_phonon_2019}}}\propto d^{-5}$ and, by combining the Langevin equations with thermal fluctuations and dissipation, leads to an analytical expression for the thermalization temperature (Eq.~(S16) of the SI in~\cite{fong_phonon_2019}). Within this framework, the interaction between the membranes is attributed to vacuum electromagnetic fluctuations through an effective Casimir-mediated coupling. Although this phenomenological description successfully reproduces the experimental observations, it does not provide a microscopic quantum derivation of the effective coupling $g_{_\text{\cite{fong_phonon_2019}}}$. This interpretation was subsequently questioned in~\cite{biehs2020fundamental,biehs2020}, who argued that the measurements in~\cite{fong_phonon_2019} do not constitute direct evidence of Casimir-force-mediated heat transfer. Instead, they proposed an alternative mechanism based on near-field radiative heat transfer mediated by thermally populated electromagnetic modes between the two solids.

A recent first-principles attempt was provided by Zhou and Liao~\cite{zhou2025microscopic}, who derived both the microscopic Hamiltonian and the effective coupling between the two mechanical modes. Although their work identifies the appropriate microscopic route, it does not quantitatively account for the observations in~\cite{fong_phonon_2019}. Three limitations are particularly relevant. First, the theory is restricted to $1$D cavity, yielding a coupling $g \propto d^{-3}$, rather than the experimentally inferred scaling $g_{_\text{\cite{fong_phonon_2019}}}\propto d^{-5}$, whose recovery requires the full $3$D geometry. Second, the finite membrane reflectivity and unequal surface areas were not included. Third, the contribution of thermally populated electromagnetic modes was neglected. This last point is central to the debate raised in~\cite{biehs2020fundamental,biehs2020}: the issue is not whether vacuum-mediated phonon interactions can exist, but whether they dominate under the conditions in~\cite{fong_phonon_2019}, or whether the observations are instead governed by near-field radiative heat transfer mediated by thermal photons. A microscopic quantum theory that incorporates and quantitatively compares both mechanisms therefore remains necessary.

The following section resolves this puzzle by developing a fully $3$D quantum microscopic theory showing that effective resonant membrane-membrane interactions mediated by electromagnetic vacuum fluctuations provide the mechanism responsible for the observed thermalization, even within the standard optomechanical regime where such interactions are generally assumed to be negligible.

{\textbf{Beyond the Standard Optomechanical Approximation}.---} To identify the microscopic interaction responsible for the observed heat transfer, we consider a $3$D optical cavity bounded by two aligned nanomechanical membranes separated by a distance $d$, which confine the quantum electromagnetic field. Small cavity-length fluctuations induced by the membrane displacements $x_i$ are characterized by the dimensionless parameters $\epsilon_i=\delta x_i/d \ll 1$. Under these conditions, as detailed in Supplementary Information (SI) Section I, the quantum Hamiltonian naturally decomposes into an unperturbed and an interaction Hamiltonian containing the higher-order optomechanical processes neglected in standard cavity optomechanics;
\begin{align}\label{Art_Total_Hamiltonian}
%\hat{H}_0=& \hbar\sum_{ n}\int d^2\bold k_\perp\omega_{ n}\,\left(\hat{a}_{\bold n}^\dag \hat{a}_{\bold n}+\delta(\bold k_\perp-\bold{k}'_\perp)\right) \nonumber \\
\hat{H}_0=& \hbar\sum_{n}^{\infty}\int d^2\mathbf{k}_\perp\omega_{n}\bigg(\hat{a}_{\mathbf{n}}^\dag \hat{a}_{\mathbf{n}} +\frac{1}{2}\bigg)+\hbar\Omega \left(\hat{b}_1^\dag\hat{b}_1+\hat{b}_2^\dag\hat{b}_2\right),\nonumber\\
%\hat{H}_{I}=&\hbar  \sum_{nm}(-1)^{n+m}\int d^2\bold k_\perp\frac{4 c^2 k_n k_m}{\sqrt{\omega_{n}\,\omega_{m}}} \hat X_{\bold n}\hat X_{\bold m}\left(\epsilon_1 \hat Q_1- \epsilon_2 \hat Q_2\right)\\
\hat{H}_{I}=& \frac{\hbar c^2}{2} \sum_{nm}^{\infty} (-1)^{n+m} \int d^2 \mathbf{k}_\perp  \frac{k_n k_m}{\sqrt{\omega_{n}\,\omega_{m}}} (\hat{a}_{\mathbf{n}} +\hat{a}_{\mathbf{n}}^\dag) (\hat{a}_{\mathbf{m}} +\hat{a}_{\mathbf{m}}^\dag) [\epsilon_1(\hat{b}_1+\hat{b}_1^\dag)- \epsilon_2 (\hat{b}_2+\hat{b}_2^\dag)].
\end{align}
Here, $\mathbf{k} \equiv(k_n,\mathbf{k}_{\perp})=(\frac{n\pi}{d},k_y,k_z)$ is the wave vector, and the dispersion relation reads $\omega_n=c\sqrt{k_n^2+k_\perp^2}$. The annihilation operators associated with the cavity and membrane modes are denoted by $\hat{a}_{\mathbf{n}}\equiv\hat{a}_{(n,\mathbf{k}_\perp)}$ and $\hat{b}_{(1,2)}$, respectively; satisfying the bosonic commutation relations $[\hat{a}_{\mathbf{n}},\hat{a}_{\mathbf{m}}^\dagger]=\delta_{nm}~\delta(\mathbf{k}_\perp-\mathbf{k}'_\perp)$, and $[\hat b_i,\hat b_j^\dagger]=\delta_{ij}$.

The microscopic origin of the observed heat transfer emerges naturally from the effective Hamiltonian obtained by applying the second-order James method~\cite{Shao2017} to the full quantum Hamiltonian \eqref{Art_Total_Hamiltonian} (see SI Section II for details). The resulting effective Hamiltonian consists of two distinct contributions: (i) an energy-shift term arising from self- and cross-Kerr interactions, which renormalizes the photonic and phononic frequencies without affecting the membrane-membrane dynamics, and (ii) the effective membrane-membrane interaction, 
\begin{align}\label{Art_Interaction_term}
\hat{H}_{\rm JI}^{(2)} =& - \hbar c^4 \epsilon_1 \epsilon_2  \sum_{n} \int d^2 \mathbf{k}_\perp \frac{k_n^4}{\omega_{n}^3} \Big (2\hat a_{\mathbf{n}}^\dagger \hat a_{\mathbf{n}} + \frac{A_\perp}{(2\pi)^2} \Big ) \Big ( \hat{b}_1^\dagger \hat{b}_2  + \hat{b}_1  \hat{b}_2^\dagger \Big ),
\end{align}
%\begin{align}\label{Interaction_term}
%\hat{H}_{\rm JI}^{(2)} =& - \hbar \sum_{n}^{\infty} g_n \Big (2\hat a_{\bold n}^\dagger \hat a_{\bold n} + \frac{A_\perp}{(2\pi)^2} \Big ) \Big ( \hat{b}_1^\dagger \hat{b}_2  + \hat{b}_1  \hat{b}_2^\dagger \Big ),
%\end{align}
%where $g_n=  c^4 \epsilon_1 \epsilon_2  \int d^2\bold k_\perp \frac{k^4_n }{\omega_{n}^3}$ \textcolor{red}{per il discorso dell'altra volta, immagino che l'integrale vada fuori insieme alla sommatoria. Giusto?} is the phonon-phonon coupling due to the $n$-th cavity mode. 
where $A_\perp$ denotes  the membrane surface area. Importantly, $\hat{H}_{\rm JI}^{(2)}$ \emph{emerges precisely in the standard optomechanical regime}, $\omega_n \gg \Omega$, \emph{where the higher-order optomechanical interactions from which it originates are conventionally neglected}. Nevertheless, these interactions generate the effective membrane-membrane coupling responsible for the observed vacuum-mediated heat transfer.
%in which direct membrane-membrane interactions are typically neglected.   
%with $\eta_n=\int d^2\bold k_\perp \omega_{n}$.
%\begin{eqnarray}
%\label{James_Hamiltonian}
%\hat H_{\rm J_{eff}}^{(2)} &=& \hat{H}_0+ \hat{H}_{\rm shift}^{(2)} + %\hat{H}_{\rm JI}^{(2)} \, , \nonumber \\
%\hat{H}_{\rm shift}^{(2)} &=&  \frac{2 \hbar \epsilon^2 \omega^2_c (4 \omega_c + %3 \Omega)}{\Omega (2 \omega_c +  \Omega)} \hat{a}^\dagger \hat{a} \nonumber \\
%&&+ \frac{2 \hbar \epsilon^2 \omega^2_c (8 \omega^2_c + 3 \Omega^2)}{\Omega (4 \omega^2_c +  \Omega^2)} \hat{a}^{\dagger 2} \hat{a}^2 \nonumber \\
%&&+ \frac{2 \hbar \epsilon^2 \omega^3_c}{4 \omega^2_c +  \Omega^2} (2\hat{a}^{\dagger} \hat{a} + \mathds{1}) (\hat{b}_1^\dagger \hat{b}_1 + \hat{b}_2^\dagger \hat{b}_2)  \nonumber  \\
%\hat{H}_{\rm JI}^{(2)} &=&  -\frac{2 \hbar \epsilon^2 \omega^3_c}{4 \omega^2_c +  \Omega^2} (2\hat{a}^{\dagger} \hat{a} + \mathds{1}) (\hat{b}_1^\dagger \hat{b}_2 + \hat{b}_1 \hat{b}_2^\dagger) \, .
%\end{eqnarray}
Furthermore, $\hat{H}_{\rm JI}^{(2)}$ can be diagonalized, allowing the Hilbert space to be decomposed into sectors with a fixed photon number. Within this representation, the effective phonon-phonon coupling naturally separates into two contributions: (i) a thermal component 
\begin{align}\label{Art_thermal_term}
g_{\rm th}= 2  \hbar c^4 \epsilon_1 \epsilon_2  \sum_{n} \int d^2 \mathbf{k}_\perp \frac{k_n^4}{\omega_{n}^3} \langle \hat{a}_{\mathbf{n}}^\dagger \hat{a}_{\mathbf{n}}\rangle,
\end{align}
originating from the thermally populated electromagnetic modes and proportional to the intracavity photon occupation, and (ii) a Casimir contribution,
\begin{align}\label{Art_Casimir_term}
g_{\rm c}= \frac{\hbar c^4 \epsilon_1 \epsilon_2 A_\perp}{(2\pi)^2}\sum_{n} \int d^2 \mathbf{k}_\perp \frac{k_n^4}{\omega_{n}^3},
\end{align}
arising from the vacuum electromagnetic field. Remarkably, these two terms provide, within a unified microscopic framework, the interaction mechanisms underlying the two competing interpretations in~\cite{fong_phonon_2019}: the thermal contribution corresponds to the near-field photonic radiative mechanism proposed in~\cite{biehs2020fundamental}, whereas the vacuum contribution reproduces the Casimir-mediated interaction invoked in~\cite{fong_phonon_2019}. As we show below, evaluating both contributions for the experimental parameters in~\cite{fong_phonon_2019} reveals that the Casimir term dominates the thermal one, thereby providing a microscopic justification for the experimental interpretation. 

Although the experiments~\cite{fong_phonon_2019} operates under high-vacuum conditions, the membranes remain near room temperature, so thermally excited mechanical motion can populate the intracavity electromagnetic field through radiative processes. The resulting photon population is well described by a Bose–Einstein distribution characterized by an effective temperature $T_{\rm eff}$, determined by the combined thermal radiation of the two membranes. Within a Stefan–Boltzmann description, $T_{\rm eff}$ is set by their respective bath temperatures, emitting areas, and emissivities, assuming each membrane remains in local thermal equilibrium with its reservoir. Consequently, the intracavity electromagnetic modes are thermally populated according to $\langle \hat a_{\mathbf{n}}^\dagger \hat a_{\mathbf{n}} \rangle = (e^{\hbar\omega_n /k_{\rm B} T_{\rm eff}}-1)^{-1}$, giving rise to the thermal contribution $g_{\rm th} = c  A_{\perp} \epsilon_1 \epsilon_2\lambda_T^{-4}  d  /75$, where $\lambda_T=\hbar c /k_{\rm B} T_{\rm eff}$ is the thermal wavelength.  By contrast, the Casimir contribution is readily found to be  $g_{\rm c}=  \pi^2  c A_\perp \epsilon_1 \epsilon_2   d^{-3} / 240$, therefore,  the ratio between the thermal and Casimir contributions scales as $g_{\rm th}/g_{\rm c} \sim (d/\lambda_{T})^4$,  showing that the competition between the two mechanisms is governed by the dimensionless parameter $d/\lambda_{T}$ (see SI Section III). According to the parameter values reported in~\cite{fong_phonon_2019} one finds $\lambda_{T} \gg d$, implying that, although the microscopic theory naturally incorporates both interaction mechanisms,  \emph{the effective membrane-membrane coupling is overwhelmingly dominated by the Casimir contribution}.

The Casimir contribution scales as $g_{\rm c} \propto d^{-5}$, rather than the usual $d^{-3}$, because the zero-point motion of the mechanical modes enters through the dimensionless displacement amplitudes $\epsilon_{(1,2)}$. This result, however, is obtained in the idealized limit of perfectly reflecting parallel plates. The experimental device in~\cite{fong_phonon_2019}, by contrast, consists of gold-coated $\rm Si_3N_4$ membranes with finite reflectivity and unequal surface areas. A quantitative comparison with the experiment therefore requires a realistic evaluation of the Casimir interaction within Lifshitz theory~\cite{lifshitz1992theory}, where finite conductivity and material dispersion are incorporated through frequency-dependent Fresnel coefficients. The resulting interaction is reduced with respect to the ideal case and can be expressed as $g(d,T) = \chi(d,T)\, g_{\rm c}(d)$. For the experimental parameters in~\cite{fong_phonon_2019}, the correction factor is accurately described by $\chi \simeq 0.045 \left(300~{\rm nm}/d\right)^{0.22}$ yielding the realistic Casimir-mediated coupling $g \simeq \pi \left(300~{\rm nm}/d \right)^{5.22}~{\rm Hz}$.
\begin{figure}[t]
    \centering
    \includegraphics[width=1\textwidth]{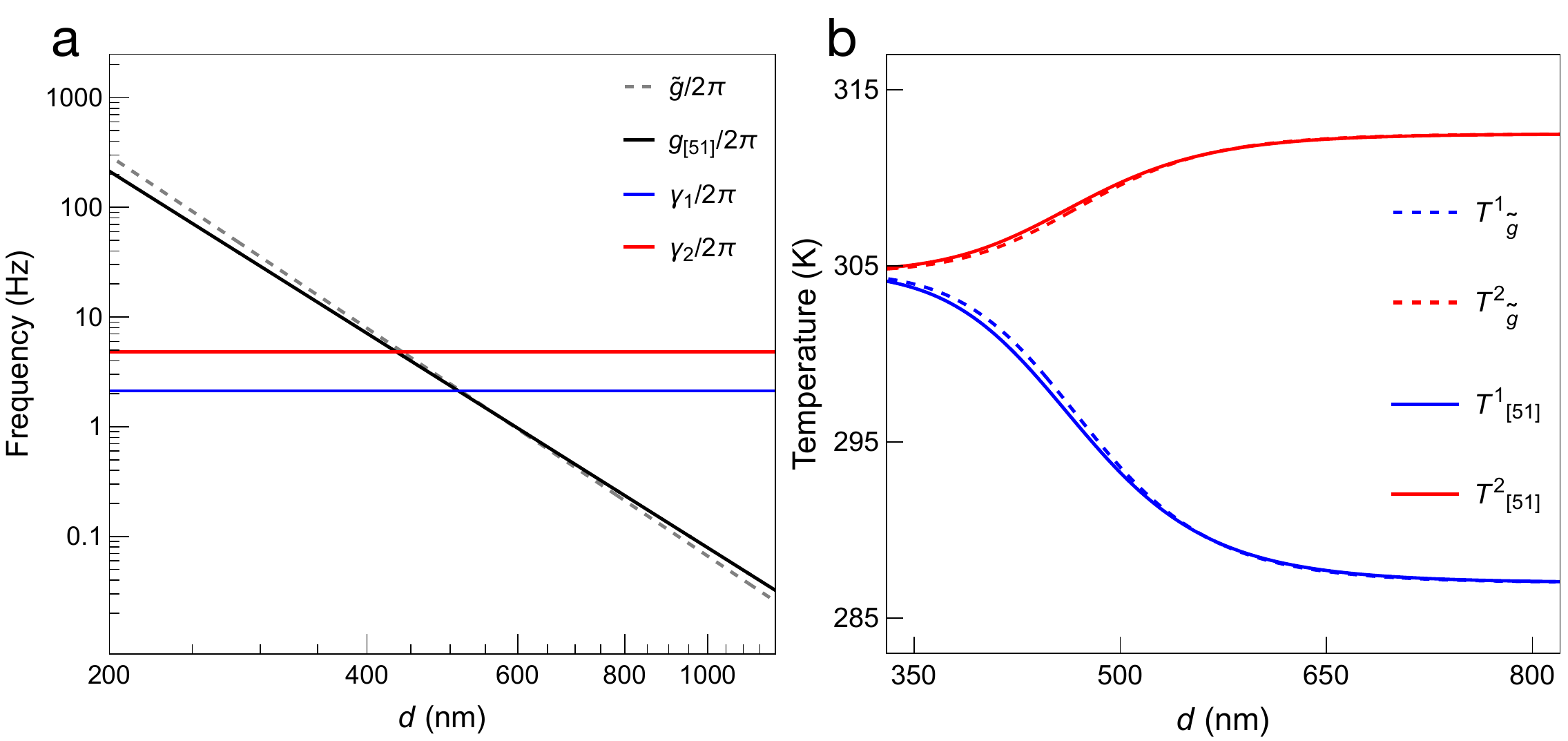}
    \caption{Comparison between the present quantum model and the experimental results of Ref.~\cite{fong_phonon_2019}.  (a) Casimir-mediated phonon coupling and mode-bath dissipation rates $\gamma_{(1,2)}=\Omega/2Q_{(1,2)}$ in logarithmic scale,  as functions of the membrane separation $d$ with $Q_2=2 \times 10^4$, $Q_1=2.25~Q_2$ taken from~\cite{fong_phonon_2019}. As $d$ decreases, the system evolves from the weak- to the strong-coupling regime. (b) Corresponding phononic thermalization as a function of $d$. The coupling extracted from Ref.~\cite{fong_phonon_2019} satisfies  $g_{_\text{\cite{fong_phonon_2019}}} \approx 2g=\Tilde{g}$, accounting for the factor of two appearing in \eqref{Art_Thermalization}.}  
    \label{fig: Comparison}
\end{figure}

Figure~\ref{fig: Comparison}\textcolor{blue}{(a)} compares the Casimir-mediated phonon coupling obtained from the present microscopic theory with the phenomenological coupling introduced in~\cite{fong_phonon_2019}, together with the mode-bath dissipation rates $\gamma_{(1,2)}$. The excellent agreement demonstrates that the phenomenological description in~\cite{fong_phonon_2019} naturally emerges from the present microscopic theory. We further find the simple relation $g_{_\text{\cite{fong_phonon_2019}}} \approx 2g=\Tilde{g}$. As shown in the next section, this relation enters directly into the thermalization dynamics, providing the microscopic connection to the phenomenological model in~\cite{fong_phonon_2019}.

{\textbf{ Expected Experimental Results from the Microscopic Theory}.---} The microscopic theory developed above predicts the experimental observations from first principles. Since the Casimir contribution dominates under the experimental conditions in~\cite{fong_phonon_2019}, the photonic degrees of freedom can be traced out, yielding the effective Hamiltonian
\begin{align}\label{Art_Final_hamiltonian}
\hat{H}_{\rm{eff}} =\hbar \Bigl(\Omega + \frac{\epsilon_1}{\epsilon_2} g  \Bigl) \Bigl(\hat{b}_1^\dag\hat{b}_1+\frac{1}{2}\Bigl)+ \hbar \Bigl(\Omega + \frac{\epsilon_2}{\epsilon_1} g \Bigl) \Bigl(\hat{b}_2^\dag\hat{b}_2+\frac{1}{2}\Bigl) - \hbar g \left(\hat{b}_{1}^\dagger \hat b_{2} + \hat{b}_{1} \hat b_{2}^\dagger \right),
\end{align}
%with $j=(2,1)$ for $i=(1,2)$ (see Supp.~Mat.~\cite{SupMat}).
Each membrane is coupled to an independent thermal reservoir, so that the dissipative dynamics is governed by the Lindblad master equation $\dot{\hat{\rho}} = -i/\hbar[\hat{H}_{\rm{eff}}, \hat{\rho}] 
+ \sum_j^2 \gamma_j \Bigl[(n^{\mathrm{th}}_j+1) \mathcal{D}[\hat{b}_j]\hat{\rho} + n^{\mathrm{th}}_j \mathcal{D}[\hat{b}^{\dagger}_j]\hat{\rho}\Bigl]$, where  \(\mathcal{D}[\hat{\bullet}]\hat{\rho} =  \hat{\bullet} \hat{\rho} \hat{\bullet}^\dagger - 1/2 \{ \hat{\bullet}^\dagger \hat{\bullet},\hat{\rho}\} \) denotes the Lindblad dissipator \cite{kurizki_kofman_2022}. The thermal phonon occupation of $j$-bath  is given by $n^{\mathrm{th}}_j = (e^{\hbar\Omega /k_{\rm B} T_j^{\rm bath}}-1)^{-1}$ where $T_j^{\rm bath}$ is the temperature of the corresponding phononic reservoir. To characterize the steady-state heat transfer mediated by quantum vacuum fluctuations, we evaluate the second-order correlation functions $B_j = \langle b_j^\dagger b_j \rangle$ and $C = \langle b_1^\dagger b_2 \rangle$ (see SI Section IV for details). Their equations of motion follow directly from the stationary condition $\rm{Tr}[\hat{O}\mathcal{L}\hat{\rho}_{ss}]=0$ whose solution yields, 
\begin{equation}\label{Art_Solution}
B_j= n_j^{\mathrm{th}} + \frac{\gamma_j (n_j^{\mathrm{th}} - n_k^{\mathrm{th}})}{(\gamma_k +  \gamma_j) \Bigl(1+\dfrac{\gamma_k \gamma_j}{4 g^2} + \Delta^2 \dfrac{ \gamma_k \gamma_j}{(\gamma_k + \gamma_j)^2} \Bigl)}.
\end{equation}
The steady-state occupations $B_j$ define, in general, an effective phonon temperature $T_j^{'}\neq T_j^{\rm{bath}}$,  arising from phonon exchange between the membranes, such that  $B_j={n'}^{\mathrm{th}}_j=(e^{\hbar\Omega /k_{\rm B} T_j^{'}}-1)^{-1}$. Both baths are kept near room temperature with $T_1^{\rm{bath}} < T_2^{\rm{bath}}$. Since for membrane frequencies of order  $10^3$ Hz $\hbar \Omega/k_{\rm B} T_{(1,2)}^{'}\ll 1$, \eqref{Art_Solution} remains valid in the high-temperature limit ${n'}^{\mathrm{th}}_{(1,2)} \simeq k_{\rm B} T_{(1,2)}^{'}/\hbar \Omega$, so that one obtains, 
\begin{equation}\label{Art_Thermalization}
T_j^{'}= T_j^{\rm{bath}} + \frac{\gamma_j (T_j^{\rm{bath}} - T_k^{\rm{bath}})}{(\gamma_k + \gamma_j) \Bigl(1+\dfrac{\gamma_k \gamma_j}{ \Tilde{g}^2} + \Delta^2 \dfrac{ \gamma_k \gamma_j}{(\gamma_k + \gamma_j)^2}\Bigl)},
\end{equation}
where $\Tilde{g}=2g$, $\Delta=(\epsilon_1^2-\epsilon_2^2)/\epsilon_1\epsilon_2$, and $k=(2,1)$ for $j=(1,2)$. Remarkably, we find that $g_{_\text{\cite{fong_phonon_2019}}} \approx 2g$, which accounts for the factor of two appearing in the thermalization formula \eqref{Art_Thermalization}.

Unlike the phenomenological model in~\cite{fong_phonon_2019}, our microscopic derivation includes the correction  $\Delta^2 \gamma_k \gamma_j/(\gamma_k + \gamma_j)^2$, accounting for the asymmetry of the membrane effective masses (see SI Section IV for details). For the experimental parameters in~\cite{fong_phonon_2019}, this correction is of order $10^{-2}$ and could therefore be neglected, \emph{recovering the thermalization formula of}~\cite{fong_phonon_2019}. In the weak-coupling regimes, $\Tilde{g} \ll \gamma_1,\gamma_2$,   \eqref{Art_Thermalization} reduces to $T_j^{'}= T_j^{\rm{bath}}$, whereas  in the strong-coupling regime $\Tilde{g} \gg \gamma_1,\gamma_2$ the membranes thermalize to a common temperature $T_1^{'}= T_2^{'}= (\gamma_1 T_1^{\rm{bath}} + \gamma_2 T_2^{\rm{bath}})/(\gamma_1+ \gamma_2)$. This behavior is shown in  \figref{fig: Comparison}\textcolor{blue}{(b)}, which highlights the agreement between the present microscopic theory and the experimental findings in~\cite{fong_phonon_2019}.

{\textbf{Phonon-phonon energy exchange drives a thermodynamic cycle}.---} Having established the microscopic mechanism responsible for the observed heat transfer, we now show that the same experimental platform can operate as a genuine quantum heat engine. By periodically modulating the membrane separation, the Casimir-mediated phonon-phonon interaction enables the system to operate as a quantum piston, driving a thermodynamic cycle that extracts net work with finite efficiency. 

To operate the system as a quantum heat engine, we introduce a time-dependent phonon-phonon coupling $g(t)=g_0+A\sin(\theta t)$, induced by a controlled harmonic modulation of the membrane separation. We consider the adiabatic regime, where the modulation frequency is much smaller than both the mechanical frequency and the dissipation rates, $\theta\ll \Omega, \gamma_{(1,2)}$, such that $ \zeta=\theta/\Gamma  \ll 1$ with  $\Gamma=\gamma_{1}+\gamma_{2}$. This introduces two well-separated timescales: a slow timescale $\tau = \zeta \Gamma t = \theta t $, associated with the external modulation, and a fast timescale $\tau/\zeta = \Gamma t$, describing the relaxation toward the instantaneous steady state. The resulting separation of timescales allows the system to continuously adapt to the slowly varying coupling, so that the instantaneous normal modes remain well defined throughout the cycle. Consequently, the effective Hamiltonian in \eqref{Art_Final_hamiltonian} can be generalized to the time-dependent form through the coupling $g(\tau)$ (see SI Section V for details). The dissipative dynamics remains governed by the same Lindblad master equation used before. 

For any time-independent system operator, the master equation gives, $\langle \dot{\hat{O}}\rangle=\rm{Tr}[\hat{O}\dot{\hat{\rho}}]=\rm{Tr}[\hat{O}\mathcal{L}(t)\hat{\rho})]$. Applying this relation to the phonon occupations $B_j = \langle b_j^\dagger b_j \rangle$ and coherence $C = \langle b_1^\dagger b_2 \rangle=R+iI$  yields a closed set of linear differential equations with coefficients determined with a  time-dependent coupling. In the adiabatic regime, $\zeta =\theta/\Gamma \ll 1$, we solve these equations using a multiple-scale expansion that separates the fast relaxation towards the instantaneous steady state from the slow evolution induced by the modulation. As detailed in SI Section V(A,B), collecting the relevant moments into the vector ${\mathbf{v}}=(B_1,B_2,I,R)^T_{t}$, and introducing the slow time $\tau=\theta t$, one obtains ${\mathbf{v}}(\tau) = {\mathbf{v}}^{(0)}(\tau) + \zeta {\mathbf{v}}^{(1)}(\tau) + \mathcal{O}(\zeta^2)$,  where ${\mathbf{v}}^{(0)}(\tau)$ describes the instantaneous quasistatic response and ${\mathbf{v}}^{(1)}(\tau)$  is the leading nonequilibrium correction induced by coupling modulation. 

To evaluate the energetic performance of the nanomechanical heat engine, we adopt the framework of quantum thermodynamics for open systems. The instantaneous internal energy is defined as $E(\tau) = \text{Tr}[H(\tau)\rho(\tau)]$, and its evolution obeys the first law of thermodynamics, $\dot{E}(\tau) = \dot{W}(\tau) + \sum_{j=1,2} \dot{Q}_j(\tau)$. Here,  $\dot{W}(\tau)$ denotes the mechanical power associated with the external modulation of the intermembrane separation, whereas  $\dot{Q}_j(\tau)$ is the heat current flowing into the system from the $j$-th thermal reservoir. When expressed in terms of the perturbative solution ${\mathbf{v}}(\tau)$, both thermodynamic currents can be written as ${\dot W}(\tau)={\dot W}^{(0)}(\tau) + \zeta {\dot W}^{(1)}(\tau) + \mathcal{O}(\zeta^2)$ and  ${\dot Q}_j(\tau)={\dot Q}_j^{(0)}(\tau) + \zeta {\dot Q}_j^{(1)}(\tau) + \mathcal{O}(\zeta^2)$, whose integration over one cycle gives the corresponding work and heat, from which the engine efficiency follows. 
%Over one modulation period $T=2\pi/\theta$, the zeroth-order contribution to the cycle work vanishes because the corresponding generalized force $\mathcal{W}(\tau)$ is a single-valued function of the instantaneous coupling, therefore, $W_{\rm cyc}^{(0)} =  \hbar  \theta \int_0^{2\pi} d\tau\, \dot{g}(\tau) \mathcal{W}[g(\tau)] = \hbar \theta  \oint dg \mathcal{W}(g)= 0$. Likewise, the first-order heat-current correction can be written as ${\dot Q}_j(\tau)= \dot{g}(\tau) \mathcal{Q}_j[g(\tau)]$, where $\mathcal{Q}_j[g]$  is single valued; its integral over a complete cycle therefore reduces to a closed line integral in $g$  and also vanishes.
As detailed in the SI Section V(C), the zeroth-order mechanical power and the first-order heat currents can be expressed as $\dot{W}^{(0)}= \dot{g}(\tau) \mathcal{W}[g(\tau)] $, and $\dot{Q}_j^{(1)}= \dot{g}(\tau) \mathcal{Q}_j[g(\tau)]$. Because $ \mathcal{W}(g)$ and $ \mathcal{Q}_j(g)$  are single-valued functions of the instantaneous coupling and  $g(T)=g(0)$ after one modulation period  $T=2\pi/\theta$, the corresponding cycle integrals reduce to closed line integrals in $g$ and therefore vanish, $W_{\rm cyc}^{(0)}=0$ and  $Q_{j, \rm cyc}^{(1)}=0$. 

Consequently, the first non-vanishing cycle work originates from the leading adiabatic lag of the system state, encoded in $\dot{W}^{(1)}(\tau)$, whereas the heat exchanged over one cycle is determined by the zeroth-order currents  $\dot{Q}^{(0)}_j(\tau)$. For the harmonic modulation $g(\tau)=g_0+A\sin(\tau)$, the corresponding closed-form expressions are derived in the Supplementary Section V. One important sign clarification: since $\dot{W}^{(1)}(\tau)>0$ denotes work performed on the system, the expression with the leading minus sign should be written for $W_{\rm cyc}$, while the extracted work is  $W_{\rm ext}=-W_{\rm cyc}$. In particular we find that, up to a positive parameter-dependent prefactor, the cycle work obeys $W_{\rm cyc}^{\rm ext} \propto -4\pi\hbar  \zeta (n_1^{\mathrm{th}}-n_2^{\mathrm{th}}) (\gamma_2^2-\gamma_1^2)\approx  -4\pi\dfrac{k_{\rm B}}{\Omega}\zeta (T_1^{\rm{bath}}-T_2^{\rm{bath}})(\gamma_2^2-\gamma_1^2)$, in the high-temperature regime $\hbar \Omega/k_{\rm B} T\ll 1$. The cycle work is therefore linear in $\zeta=\theta/\Gamma$, as expected for the leading finite-rate adiabatic response, while its sign is jointly controlled by the thermal gradient and the asymmetry between the dissipation rates. Therefore, the net work extraction requires $(T_1^{\rm{bath}}-T_2^{\rm{bath}}) (\gamma_2^2-\gamma_1^2) >0 $. Assuming a negative external thermal gradient ($T_1^{\rm{bath}} < T_2^{\rm{bath}}$), the system functions as an quantum engine if $\gamma_2<\gamma_1$. Although the experiment of Ref.~\cite{fong_phonon_2019} employed the opposite condition $\gamma_2>\gamma_1$, the same platform could access the engine regime by engineering $\gamma_2<\gamma_1$.
\begin{figure*}[t]
    \centering
    \includegraphics[width=1\textwidth]{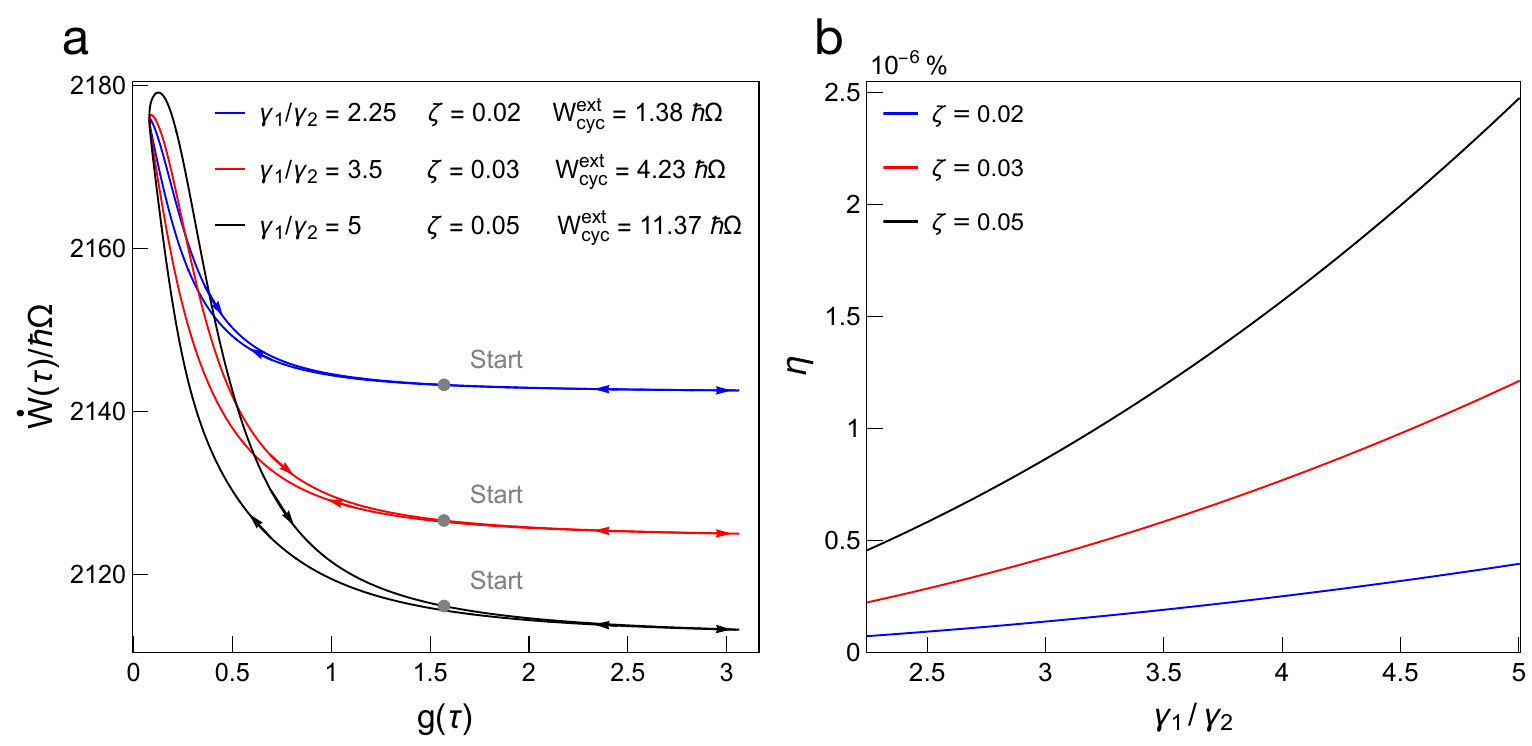}
    \caption{Hysteresis loop $(g(\tau), \dot{W}(\tau)/\hbar \Omega)$ plotted in (a) for different values of the adiabatic parameter $\zeta=\theta/\Gamma$ and the losses-ratio $\gamma_1/\gamma_2$. The nonzero area is related to net-work $W_{\rm cyc}^{\rm ext}/ \hbar \Omega$ extracted from a cycle under the slow-time modulation $g(\tau)=g_0+A\sin(\tau)$, with $g_0= 1.57~\rm{Hz}$, and $A= 1.49~\rm{Hz}$. The slow-time modulation $g(\tau)$ bring the system from weak- to strong-coupling regime over a cycle at the rate $\tau=\theta t$. In (b) the efficiency as function of the losses ratio $\gamma_1/\gamma_2$, and for different value of the adiabatic parameter $\zeta=\theta/\Gamma$. Efficiency increases when the loss asymmetry does, as well as increasing $\zeta$,  which makes faster the coupling-modulation. The other system parameters are taken from  Ref.~\cite{fong_phonon_2019}, $\Delta=0.31,~\Omega/2\pi= 191.6 \times 10^{3}~\rm{Hz},~T_{1}^{\rm{bath}}=287~K, \rm{and}~T_{2}^{\rm{bath}}=312.5~K$.}    
    \label{fig: Hysteresis_loop}
\end{figure*}
In \figref{fig: Hysteresis_loop}\textcolor{blue}{(a)}, we show the hysteresis loops traced by $(g(\tau), \dot{W}(\tau)/\hbar \Omega)$, for different value of the adiabatic parameter $\zeta=\theta/\Gamma$ and the losses-ratio $\gamma_1/\gamma_2$. The remaining parameters $(\Delta, \Omega, T_{(1,2)}^{\rm{bath}})$ are taken from Ref.~\cite{fong_phonon_2019}. Since the zeroth-order power gives no net contribution when integrated over a closed cycle, the oriented finite area enclosed by the loops reflects the principal non-equilibrium response and measures the non-zero work extracted.  $W_{\rm cyc}^{\rm ext}/ \hbar \Omega$. The harmonic modulation  $g(\tau)=g_0+A\sin(\tau)$ drives the system from the weak- to the strong-coupling regime and back, while the finite relaxation lag generates the hysteretic response underlying quantum heat-engine operation.

Because the first-order heat contribution vanishes over a complete cycle, the leading-order efficiency is given by the ratio between the extracted work and the zeroth-order heat absorbed from the second reservoir, $\eta=W_{\rm cyc}^{\rm ext}/Q_{2,{\rm cyc}}^{(0)}$. In  \figref{fig: Hysteresis_loop}\textcolor{blue}{(b)},  we plot the thermal efficiency versus the dissipation-rate ratio $\gamma_1/\gamma_2$ for three values of the adiabatic parameter $\zeta$, which controls the coupling-modulation rate. Over the range shown, the efficiency $\eta$ increases with both the dissipation asymmetry and $\zeta$. This behaviour reflects the enhancement of the leading finite-time work contribution by asymmetric dissipation and faster driving, while remaining within the adiabatic regime.

{\textbf{Discussion}.---} We have established the microscopic quantum mechanism underlying the heat transfer observed between the spatially separated membranes in~\cite{fong_phonon_2019}, resolving the ambiguity surrounding the physical origin of their experimental findings. Starting from the full mirror-field Hamiltonian, we have shown that the higher-order optomechanical interactions conventionally neglected in the standard regime generate an effective phonon-phonon coupling that cannot be captured by the radiation-pressure interaction alone. The resulting microscopic theory identifies both thermal and vacuum contributions to the membrane coupling and demonstrates that, under the experimental conditions, the latter dominates. The energy exchange is therefore mediated by quantum fluctuations of the electromagnetic vacuum confined between the membranes. When realistic material properties are incorporated through Lifshitz theory, this mechanism quantitatively accounts for the measured coupling strength and distance dependence, the crossover from weak to strong coupling, and the observed phonon thermalization. These results provide the missing microscopic description of the experiments and resolve the debate concerning the origin of the observed heat transfer.
Having established this microscopic mechanism, we have further shown that the same platform can support a genuine thermodynamic cycle. Periodically modulating the membrane separation controls the vacuum-mediated phonon-phonon interaction and drives the system between the weak- and strong-coupling regimes. The resulting finite-time lag of the mechanical state, combined with asymmetric dissipation, generates a hysteretic response that enables net work extraction at finite efficiency. The thermal bias supplies the extracted energy, while vacuum fluctuations provide the microscopic channel through which energy is transferred and converted. More broadly, our findings demonstrate that higher-order mirror-field interactions are not merely negligible corrections to linearized optomechanics, but can determine both the transport and thermodynamic properties of experimentally accessible devices. They thus provide a microscopic framework for exploring the next generation of optomechanical phenomena beyond the radiation-pressure approximation.

{\textbf{Acknowledgments}.---}
\noindent The authors thank Vincenzo Savona and Alberto Mercurio for fruitful discussions. V.M. acknowledge PNRR MUR project “National Quantum Science and Technology Institute” – NQSTI (Grant No. PE0000023). F.N. is supported in part by the Japan Science and Technology Agency (JST) [via the CREST Quantum Frontiers program Grant No. JPMJCR24I2, the Quantum Leap Flagship Program (Q-LEAP), the Moonshot R$\&$D Grant Number JPMJMS256E, and the ASPIRE program (Grant Number JPMJAP2513)]. A.F. acknowledges the ``FY2026 JSPS Postdoctoral Fellowships for Research in Japan (Standard)'', sponsored by the Japanese Society for the Promotion of Science (JSPS).

\bibliography{alessandria}

@article{biehs2020,
  title={On the heat transfer across a vacuum gap mediated by Casimir force},
  author={Biehs, Svend-Age and Kittel, Achim and Ben-Abdallah, Philippe},
  journal={arXiv preprint arXiv:2003.00760},
  year={2020},
  url = {https://arxiv.org/abs/2003.00760}
}

@article{Degen2017,
  title = {Quantum sensing},
  author = {Degen, C. L. and Reinhard, F. and Cappellaro, P.},
  journal = {Rev. Mod. Phys.},
  volume = {89},
  issue = {3},
  pages = {035002},
  numpages = {39},
  year = {2017},
  month = {Jul},
  publisher = {American Physical Society},
  doi = {10.1103/RevModPhys.89.035002},
  url = {https://link.aps.org/doi/10.1103/RevModPhys.89.035002}
}

@book{peskin2018,
  title={An Introduction to quantum field theory},
  author={Peskin, Michael E},
  year={2018},
  publisher={CRC press}
}

@article{bragadin2026,
  title={Dissipative phase transitions and chaos in two-photon driven quantum optomechanics},
  author={Bragadin, Giovanni and Ferrari, Filippo and Fioroni, Lorenzo and Macr{\`\i}, Vincenzo and Savona, Vincenzo and Mercurio, Alberto},
  journal={arXiv preprint arXiv:2607.12044},
  year={2026},
  url = {https://arxiv.org/abs/2607.12044}
}

@article{biehs2020fundamental,
  title={Fundamental limitations of the mode temperature concept in strongly coupled systems},
  author={Biehs, Svend-Age and Kittel, Achim and Ben-Abdallah, Philippe},
  journal={Z. Naturforsch. A},
  volume={75},
  number={9},
  pages={803--807},
  year={2020},
  publisher={De Gruyter},
  doi = {10.1515/zna-2020-0204},
}

@book{milton2001,
  title={The Casimir effect: physical manifestations of zero-point energy},
  author={Milton, Kimball A},
  year={2001},
  publisher={World Scientific}
}

@incollection{lifshitz1992theory,
  title={The theory of molecular attractive forces between solids},
  author={Lifshitz, Evgenni Mikhailovich and Hamermesh, M and others},
  booktitle={Perspectives in theoretical physics},
  pages={329--349},
  year={1992},
  publisher={Elsevier},
  url = {https://www.sciencedirect.com/science/chapter/edited-volume/abs/pii/B9780080363646500314}
}

@book{kurizki_kofman_2022, place={Cambridge}, title={Thermodynamics and Control of Open Quantum Systems}, publisher={Cambridge University Press}, author={Kurizki, Gershon and Kofman, Abraham G.}, year={2022}
}

@article{weis2010,
  title={Optomechanically induced transparency},
  author={Weis, Stefan and Rivi{\`e}re, R{\'e}mi and Del{\'e}glise, Samuel and Gavartin, Emanuel and Arcizet, Olivier and Schliesser, Albert and Kippenberg, Tobias J},
  journal={Science},
  volume={330},
  number={6010},
  pages={1520--1523},
  year={2010},
  publisher={American Association for the Advancement of Science},
  url = {https://www.science.org/doi/full/10.1126/science.1195596?casa_token=ymH0V_DcicEAAAAA%3A5pY3VymhxkNRwizabaeEQsfD1rUK20VFX_P-rJRI9eSpgQnkBxNwQ__F0Dp84RvV0xY08cZK_Axt}
}

@article{Kronwald2013,
  title = {Optomechanically Induced Transparency in the Nonlinear Quantum Regime},
  author = {Kronwald, Andreas and Marquardt, Florian},
  journal = {Phys. Rev. Lett.},
  volume = {111},
  issue = {13},
  pages = {133601},
  numpages = {5},
  year = {2013},
  month = {Sep},
  publisher = {American Physical Society},
  doi = {10.1103/PhysRevLett.111.133601},
  url = {https://link.aps.org/doi/10.1103/PhysRevLett.111.133601}
}

@article{Romero2011,
  title = {Large Quantum Superpositions and Interference of Massive Nanometer-Sized Objects},
  author = {Romero-Isart, O. and Pflanzer, A. C. and Blaser, F. and Kaltenbaek, R. and Kiesel, N. and Aspelmeyer, M. and Cirac, J. I.},
  journal = {Phys. Rev. Lett.},
  volume = {107},
  issue = {2},
  pages = {020405},
  numpages = {4},
  year = {2011},
  month = {Jul},
  publisher = {American Physical Society},
  doi = {10.1103/PhysRevLett.107.020405},
  url = {https://link.aps.org/doi/10.1103/PhysRevLett.107.020405}
}

@article{kippenberg2008,
  title={Cavity optomechanics: back-action at the mesoscale},
  author={Kippenberg, Tobias J and Vahala, Kerry J},
  journal={Science},
  volume={321},
  number={5893},
  pages={1172--1176},
  year={2008},
  publisher={American Association for the Advancement of Science},
  url = {https://www.science.org/doi/full/10.1126/science.1156032?casa_token=_NZH_4PwnZkAAAAA%3Af4eTPIxPUUpWyDP5a1rmpJritOvROhW0vqXPsKdslfSgaTupfKpyFJ_wtJUhYnArDbiDWIKzKuM9}
}

@article{Verlot2010,
  title = {Backaction Amplification and Quantum Limits in Optomechanical Measurements},
  author = {Verlot, P. and Tavernarakis, A. and Briant, T. and Cohadon, P.-F. and Heidmann, A.},
  journal = {Phys. Rev. Lett.},
  volume = {104},
  issue = {13},
  pages = {133602},
  numpages = {4},
  year = {2010},
  month = {Mar},
  publisher = {American Physical Society},
  doi = {10.1103/PhysRevLett.104.133602},
  url = {https://link.aps.org/doi/10.1103/PhysRevLett.104.133602}
}

@article{safavi2011,
  title={Electromagnetically induced transparency and slow light with optomechanics},
  author={Safavi-Naeini, Amir H and Alegre, TP Mayer and Chan, Jasper and Eichenfield, Matt and Winger, Martin and Lin, Qiang and Hill, Jeff T and Chang, Darrick E and Painter, Oskar},
  journal={Nature},
  volume={472},
  number={7341},
  pages={69--73},
  year={2011},
  publisher={Nature Publishing Group UK London},
  url = {https://www.nature.com/articles/nature09933}
}

@Article{Wilson2011,
  Title= {Observation of the dynamical {C}asimir effect in a superconducting circuit},
  Author                   = {Wilson, C.M. and Johansson, G. and Pourkabirian, A. and Simoen, M. and Johansson, J.R. and Duty, T. and Nori, F. and Delsing, P.},
  Journal                  = {Nature},
  Year                     = {2011},
  Number                   = {7373},
  Pages                    = {376},
  Volume                   = {479},
  Publisher                = {Nature Publishing Group},
  Url                    = {http://www.nature.com/nature/journal/v479/n7373/abs/nature10561.html}
}

@article{Shao2017,
  title = {Generalized James' effective Hamiltonian method},
  author = {Shao, Wenjun and Wu, Chunfeng and Feng, Xun-Li},
  journal = {Phys. Rev. A},
  volume = {95},
  issue = {3},
  pages = {032124},
  numpages = {4},
  year = {2017},
  month = {Mar},
  publisher = {American Physical Society},
  doi = {10.1103/PhysRevA.95.032124},
  url = {https://link.aps.org/doi/10.1103/PhysRevA.95.032124}
}

@Article{Mercurio2025,
	title={{Bilateral photon emission from a vibrating mirror and multiphoton entanglement generation}},
	author={Alberto Mercurio and Enrico Russo and Fabio Mauceri and Salvatore Savasta and Franco Nori and Vincenzo Macrì and Rosario Lo Franco},
	journal={SciPost Phys.},
	volume={18},
	pages={067},
	year={2025},
	publisher={SciPost},
	doi={10.21468/SciPostPhys.18.2.067},
	url={https://scipost.org/10.21468/SciPostPhys.18.2.067},
}

@article{Wang2023,
  title = {Coherent resonant coupling between atoms and a mechanical oscillator mediated by cavity-vacuum fluctuations},
  author = {Wang, Bo and Hu, Jia-Ming and Macr\`{\i}, Vincenzo and Xiang, Ze-Liang and Nori, Franco},
  journal = {Phys. Rev. Res.},
  volume = {5},
  issue = {1},
  pages = {013075},
  numpages = {19},
  year = {2023},
  month = {Feb},
  publisher = {American Physical Society},
  doi = {10.1103/PhysRevResearch.5.013075},
  url = {https://link.aps.org/doi/10.1103/PhysRevResearch.5.013075}
}

@article{macri2018,
	title = {Nonperturbative {Dynamical} {Casimir} {Effect} in {Optomechanical} {Systems}: {Vacuum} {Casimir}-{Rabi} {Splittings}},
	volume = {8},
	issn = {2160-3308},
	shorttitle = {Nonperturbative {Dynamical} {Casimir} {Effect} in {Optomechanical} {Systems}},
	url = {https://link.aps.org/doi/10.1103/PhysRevX.8.011031},
	doi = {10.1103/PhysRevX.8.011031},
	number = {1},
	urldate = {2021-11-21},
	journal = {Phys. Rev. X},
	author = {Macrì, Vincenzo and Ridolfo, Alessandro and Di Stefano, Omar and Kockum, Anton Frisk and Nori, Franco and Savasta, Salvatore},
	month = feb,
	year = {2018},
	pages = {011031}
}

@article{DiStefano2019,
  title = {Interaction of Mechanical Oscillators Mediated by the Exchange of Virtual Photon Pairs},
  author = {Di Stefano, Omar and Settineri, Alessio and Macr\`{\i}, Vincenzo and Ridolfo, Alessandro and Stassi, Roberto and Kockum, Anton Frisk and Savasta, Salvatore and Nori, Franco},
  journal = {Phys. Rev. Lett.},
  volume = {122},
  issue = {3},
  pages = {030402},
  numpages = {7},
  year = {2019},
  month = {Jan},
  publisher = {American Physical Society},
  doi = {10.1103/PhysRevLett.122.030402},
  url = {https://link.aps.org/doi/10.1103/PhysRevLett.122.030402}
}

@article{Ferreri2022a,
  title = {Interplay between optomechanics and the dynamical Casimir effect},
  author = {Ferreri, Alessandro and Pfeifer, Hannes and Wilhelm, Frank K. and Hofferberth, Sebastian and Bruschi, David Edward},
  journal = {Phys. Rev. A},
  volume = {106},
  issue = {3},
  pages = {033502},
  numpages = {17},
  year = {2022},
  month = {Sep},
  publisher = {American Physical Society},
  doi = {10.1103/PhysRevA.106.033502},
  url = {https://link.aps.org/doi/10.1103/PhysRevA.106.033502}
}

@article{Ferreri2024,
  title = {Phonon-photon conversion as mechanism for cooling and coherence transfer},
  author = {Ferreri, Alessandro and Bruschi, David Edward and Wilhelm, Frank K. and Nori, Franco and Macr\`{\i}, Vincenzo},
  journal = {Phys. Rev. Res.},
  volume = {6},
  issue = {2},
  pages = {023320},
  numpages = {14},
  year = {2024},
  month = {Jun},
  publisher = {American Physical Society},
  doi = {10.1103/PhysRevResearch.6.023320},
  url = {https://link.aps.org/doi/10.1103/PhysRevResearch.6.023320}
}

@article{Settineri2019,
  title = {Conversion of mechanical noise into correlated photon pairs: Dynamical Casimir effect from an incoherent mechanical drive},
  author = {Settineri, Alessio and Macr\`{\i}, Vincenzo and Garziano, Luigi and Di Stefano, Omar and Nori, Franco and Savasta, Salvatore},
  journal = {Phys. Rev. A},
  volume = {100},
  issue = {2},
  pages = {022501},
  numpages = {10},
  year = {2019},
  month = {Aug},
  publisher = {American Physical Society},
  doi = {10.1103/PhysRevA.100.022501},
  url = {https://link.aps.org/doi/10.1103/PhysRevA.100.022501}
}

@article{Aspelmeyer2014,
  title = {Cavity optomechanics},
  author = {Aspelmeyer, Markus and Kippenberg, Tobias J. and Marquardt, Florian},
  journal = {Rev. Mod. Phys.},
  volume = {86},
  issue = {4},
  pages = {1391},
  numpages = {62},
  year = {2014},
  month = {Dec},
  publisher = {American Physical Society},
  doi = {10.1103/RevModPhys.86.1391},
  url = {https://link.aps.org/doi/10.1103/RevModPhys.86.1391}
}

@article{pirkkalainen2013,
  title={Hybrid circuit cavity quantum electrodynamics with a micromechanical resonator},
  author={Pirkkalainen, J-M and Cho, SU and Li, Jian and Paraoanu, GS and Hakonen, PJ and Sillanpää, MA},
  journal={Nature},
  volume={494},
  number={7436},
  pages={211--215},
  year={2013},
  publisher={Nature Publishing Group UK London},
  url = {https://www.nature.com/articles/nature11821}
}

@article{Heikkil2014,
  title = {Enhancing Optomechanical Coupling via the Josephson Effect},
  author = {Heikkilä, T. T. and Massel, F. and Tuorila, J. and Khan, R. and Sillanpää, M. A.},
  journal = {Phys. Rev. Lett.},
  volume = {112},
  issue = {20},
  pages = {203603},
  numpages = {6},
  year = {2014},
  month = {May},
  publisher = {American Physical Society},
  doi = {10.1103/PhysRevLett.112.203603},
  url = {https://link.aps.org/doi/10.1103/PhysRevLett.112.203603}
}

@article{Nation2012,
  title = {Colloquium: Stimulating uncertainty: Amplifying the quantum vacuum with superconducting circuits},
  author = {Nation, P. D. and Johansson, J. R. and Blencowe, M. P. and Nori, Franco},
  journal = {Rev. Mod. Phys.},
  volume = {84},
  issue = {1},
  pages = {1--24},
  numpages = {0},
  year = {2012},
  month = {Jan},
  publisher = {American Physical Society},
  doi = {10.1103/RevModPhys.84.1},
  url = {https://link.aps.org/doi/10.1103/RevModPhys.84.1}
}

@article{Johansson2009,
  title = {Dynamical Casimir Effect in a Superconducting Coplanar Waveguide},
  author = {Johansson, J. R. and Johansson, G. and Wilson, C. M. and Nori, Franco},
  journal = {Phys. Rev. Lett.},
  volume = {103},
  issue = {14},
  pages = {147003},
  numpages = {4},
  year = {2009},
  month = {Sep},
  publisher = {American Physical Society},
  doi = {10.1103/PhysRevLett.103.147003},
  url = {https://link.aps.org/doi/10.1103/PhysRevLett.103.147003}
}

@article{rimberg2014,
  title={A cavity-Cooper pair transistor scheme for investigating quantum optomechanics in the ultra-strong coupling regime},
  author={Rimberg, AJ and Blencowe, MP and Armour, AD and Nation, PD},
  journal={New J. Phys.},
  volume={16},
  number={5},
  pages={055008},
  year={2014},
  publisher={IOP Publishing},
  doi =  {10.1088/1367-2630/16/5/055008}
}

@article{Nunnenkamp2011,
  title = {Single-Photon Optomechanics},
  author = {Nunnenkamp, A. and B\o{}rkje, K. and Girvin, S. M.},
  journal = {Phys. Rev. Lett.},
  volume = {107},
  issue = {6},
  pages = {063602},
  numpages = {5},
  year = {2011},
  month = {Aug},
  publisher = {American Physical Society},
  doi = {10.1103/PhysRevLett.107.063602},
  url = {https://link.aps.org/doi/10.1103/PhysRevLett.107.063602}
}

@article{girvin2florian2009,
  title={Optomechanics},
  author={Girvin, S M and  Marquardt, Florian},
  journal={Physics},
  year = {2009},
  volume={2},
  pages={40},
  publisher={APS},
  url = {https://physics.aps.org/articles/v2/40}
}

@article{mari_quantum_2015,
	title = {Quantum optomechanical piston engines powered by heat},
	volume = {48},
	issn = {0953-4075, 1361-6455},
	url = {https://iopscience.iop.org/article/10.1088/0953-4075/48/17/175501},
	doi = {10.1088/0953-4075/48/17/175501},
	number = {17},
	urldate = {2022-03-18},
	journal = {J. Phys. B-At. Mol. Opt.},
	author = {Mari, A and Farace, A and Giovannetti, V},
	month = sep,
	year = {2015},
	pages = {175501}
}

@article{zhang_quantum_2014,
	title = {Quantum {Optomechanical} {Heat} {Engine}},
	volume = {112},
	issn = {0031-9007, 1079-7114},
	url = {https://link.aps.org/doi/10.1103/PhysRevLett.112.150602},
	doi = {10.1103/PhysRevLett.112.150602},
	number = {15},
	urldate = {2022-03-18},
	journal = {Phys. Rev. Lett.},
	author = {Zhang, Keye and Bariani, Francesco and Meystre, Pierre},
	month = apr,
	year = {2014},
	pages = {150602}
}

@article{fong_phonon_2019,
	title = {Phonon heat transfer across a vacuum through quantum fluctuations},
	volume = {576},
	issn = {0028-0836, 1476-4687},
	url = {http://www.nature.com/articles/s41586-019-1800-4},
	doi = {10.1038/s41586-019-1800-4},
	number = {7786},
	urldate = {2023-04-13},
	journal = {Nature},
	author = {Fong, King Yan and Li, Hao-Kun and Zhao, Rongkuo and Yang, Sui and Wang, Yuan and Zhang, Xiang},
	month = dec,
	year = {2019},
	pages = {243--247}
 }

@article{renninger2018,
  title = {Bulk Crystalline Optomechanics},
  author = {Renninger, W. H. and Kharel, P. and Behunin, R. O. and Rakich, P. T.},
  year = {2018},
  month = jun,
  journal = {Nat. Phys.},
  volume = {14},
  number = {6},
  pages = {601--607},
  issn = {1745-2473, 1745-2481},
  doi = {10.1038/s41567-018-0090-3}
}

@article{Diamandi2025,
  title = {Optomechanical Control of Long-Lived Bulk Acoustic Phonons in the Quantum Regime},
  author = {Diamandi, Hilel Hagai and Luo, Yizhi and Mason, David and Kanmaz, Tevfik Bulent and Ghosh, Sayan and Pavlovich, Margaret and Yoon, Taekwan and Behunin, Ryan and Puri, Shruti and Harris, Jack G. E. and Rakich, Peter T.},
  year = {2025},
  month = aug,
  journal = {Nat. Phys.},
  issn = {1745-2473, 1745-2481},
  url = {https://www.nature.com/articles/s41567-025-02989-4}
}

@article{gil2015,
  title={High-frequency nano-optomechanical disk resonators in liquids},
  author={Gil-Santos, Eduardo and Baker, Christophe and Nguyen, DT and Hease, W and Gomez, Carmen and Lema{\^\i}tre, Aristide and Ducci, Sara and Leo, Giuseppe and Favero, Iv{\'a}n},
  journal={Nat. Nanotechnol.},
  volume={10},
  number={9},
  pages={810--816},
  year={2015},
  publisher={Nature Publishing Group UK London},
  url = {https://www.nature.com/articles/nnano.2015.160}
}

@article{Ding2010,
  title = {High Frequency GaAs Nano-Optomechanical Disk Resonator},
  author = {Ding, Lu and Baker, Christophe and Senellart, Pascale and Lemaitre, Aristide and Ducci, Sara and Leo, Giuseppe and Favero, Ivan},
  journal = {Phys. Rev. Lett.},
  volume = {105},
  issue = {26},
  pages = {263903},
  numpages = {4},
  year = {2010},
  month = {Dec},
  publisher = {American Physical Society},
  doi = {10.1103/PhysRevLett.105.263903},
  url = {https://link.aps.org/doi/10.1103/PhysRevLett.105.263903}
}

@article{jiang2012,
  title={High-frequency silicon optomechanical oscillator with an ultralow threshold},
  author={Jiang, Wei C and Lu, Xiyuan and Zhang, Jidong and Lin, Qiang},
  journal={Opt. Express},
  volume={20},
  number={14},
  pages={15991--15996},
  year={2012},
  publisher={Optical Society of America},
  url = {https://opg.optica.org/oe/fulltext.cfm?uri=oe-20-14-15991}
}

@article{Connell2010quantum,
  title={Quantum ground state and single-phonon control of a mechanical resonator},
  author={O’Connell, Aaron D and Hofheinz, Max and Ansmann, Markus and Bialczak, Radoslaw C and Lenander, Mike and Lucero, Erik and Neeley, Matthew and Sank, Daniel and Wang, H and Weides, Martin and others},
  journal={Nature},
  volume={464},
  number={7289},
  pages={697--703},
  year={2010},
  url = {https://www.nature.com/articles/nature08967.},
  publisher={Nature Publishing Group}
}

@article{brunelli2015,
	title = {Out-of-equilibrium thermodynamics of quantum optomechanical systems},
	volume = {17},
	issn = {1367-2630},
	url = {https://iopscience.iop.org/article/10.1088/1367-2630/17/3/035016},
	doi = {10.1088/1367-2630/17/3/035016},
	number = {3},
	urldate = {2022-03-18},
	journal = {New J. Phys.},
	author = {Brunelli, M and Xuereb, A and Ferraro, A and Chiara, G De and Kiesel, N and Paternostro, M},
	month = mar,
	year = {2015},
	pages = {035016}
}

@article{law_interaction_1995,
	title = {Interaction between a moving mirror and radiation pressure: {A} {Hamiltonian} formulation},
	volume = {51},
	issn = {1050-2947, 1094-1622},
	shorttitle = {Interaction between a moving mirror and radiation pressure},
	url = {https://link.aps.org/doi/10.1103/PhysRevA.51.2537},
	doi = {10.1103/PhysRevA.51.2537},
	number = {3},
	urldate = {2021-11-21},
	journal = {Phys. Rev. A},
	author = {Law, C. K.},
	month = mar,
	year = {1995},
	pages = {2537--2541}
}

@article{liu2021,
  title={Progress of optomechanical micro/nano sensors: a review},
  author={Liu, Xinmiao and Liu, Weixin and Ren, Zhihao and Ma, Yiming and Dong, Bowei and Zhou, Guangya and Lee, Chengkuo},
  journal={Int. J. Optomechatronics},
  volume={15},
  number={1},
  pages={120--159},
  year={2021},
  publisher={Taylor \& Francis},
	url = {https://www.tandfonline.com/doi/full/10.1080/15599612.2021.1986612},
}

@article{barzanjeh2022,
  title={Optomechanics for quantum technologies},
  author={Barzanjeh, Shabir and Xuereb, Andr{\'e} and Gr{\"o}blacher, Simon and Paternostro, Mauro and Regal, Cindy A and Weig, Eva M},
  journal={Nat. Phys.},
  volume={18},
  number={1},
  pages={15--24},
  year={2022},
  publisher={Nature Publishing Group UK London},
	url = {https://www.nature.com/articles/s41567-021-01402-0},
}

@article{teufel2011,
  title={Sideband cooling of micromechanical motion to the quantum ground state},
  author={Teufel, John D and Donner, Tobias and Li, Dale and Harlow, Jennifer W and Allman, MS and Cicak, Katarina and Sirois, Adam J and Whittaker, Jed D and Lehnert, Konrad W and Simmonds, Raymond W},
  journal={Nature},
  volume={475},
  number={7356},
  pages={359--363},
  year={2011},
  publisher={Nature Publishing Group UK London},
  url = {https://www.nature.com/articles/nature10261}
}

@article{schliesser2009,
  title={Resolved-sideband cooling and position measurement of a micromechanical oscillator close to the Heisenberg uncertainty limit},
  author={Schliesser, Albert and Arcizet, Olivier and Rivi{\`e}re, R{\'e}mi and Anetsberger, Georg and Kippenberg, Tobias J},
  journal={Nat. Phys.},
  volume={5},
  number={7},
  pages={509--514},
  year={2009},
  publisher={Nature Publishing Group UK London},
  url = {https://www.nature.com/articles/nphys1304}
}

@article{groblacher2009,
  title={Demonstration of an ultracold micro-optomechanical oscillator in a cryogenic cavity},
  author={Gr{\"o}blacher, Simon and Hertzberg, Jared B and Vanner, Michael R and Cole, Garrett D and Gigan, Sylvain and Schwab, KC and Aspelmeyer, Markus},
  journal={Nat. Phys.},
  volume={5},
  number={7},
  pages={485--488},
  year={2009},
  publisher={Nature Publishing Group UK London},
  url = {https://www.nature.com/articles/nphys1301}
}

@article{Butera2019,
  title = {Mechanical backreaction effect of the dynamical Casimir emission},
  author = {Butera, Salvatore and Carusotto, Iacopo},
  journal = {Phys. Rev. A},
  volume = {99},
  issue = {5},
  pages = {053815},
  numpages = {17},
  year = {2019},
  month = {May},
  publisher = {American Physical Society},
  doi = {10.1103/PhysRevA.99.053815},
  url = {https://link.aps.org/doi/10.1103/PhysRevA.99.053815}
}

@article{Montalbano2023,
  title = {Spatial correlations of field observables in two half-spaces separated by a movable perfect mirror},
  author = {Montalbano, Federico and Armata, Federico and Rizzuto, Lucia and Passante, Roberto},
  journal = {Phys. Rev. D},
  volume = {107},
  issue = {5},
  pages = {056007},
  numpages = {10},
  year = {2023},
  month = {Mar},
  publisher = {American Physical Society},
  doi = {10.1103/PhysRevD.107.056007},
  url = {https://link.aps.org/doi/10.1103/PhysRevD.107.056007}
}

@article{Macri2016,
  title = {Deterministic synthesis of mechanical NOON states in ultrastrong optomechanics},
  author = {Macr\'{\i}, V. and Garziano, L. and Ridolfo, A. and Di Stefano, O. and Savasta, S.},
  journal = {Phys. Rev. A},
  volume = {94},
  issue = {1},
  pages = {013817},
  numpages = {11},
  year = {2016},
  month = {Jul},
  publisher = {American Physical Society},
  doi = {10.1103/PhysRevA.94.013817},
  url = {https://link.aps.org/doi/10.1103/PhysRevA.94.013817}
}

@article{Butera2025,
  title = {Corrections to the optomechanical Hamiltonian from quadratic fluctuations of a moving mirror},
  author = {Butera, Salvatore},
  journal = {Phys. Rev. A},
  volume = {111},
  issue = {4},
  pages = {043524},
  numpages = {10},
  year = {2025},
  month = {Apr},
  publisher = {American Physical Society},
  doi = {10.1103/PhysRevA.111.043524},
  url = {https://link.aps.org/doi/10.1103/PhysRevA.111.043524}
}

@article{Butera22,
  title = {Influence functional for two mirrors interacting via radiation pressure},
  author = {Butera, Salvatore},
  journal = {Phys. Rev. D},
  volume = {105},
  issue = {1},
  pages = {016023},
  numpages = {20},
  year = {2022},
  month = {Jan},
  publisher = {American Physical Society},
  doi = {10.1103/PhysRevD.105.016023},
  url = {https://link.aps.org/doi/10.1103/PhysRevD.105.016023}
}

@article{Holmes2020,
  title = {Enhanced Energy Transfer to an Optomechanical Piston from Indistinguishable Photons},
  author = {Holmes, Zo\"e and Anders, Janet and Mintert, Florian},
  journal = {Phys. Rev. Lett.},
  volume = {124},
  issue = {21},
  pages = {210601},
  numpages = {6},
  year = {2020},
  month = {May},
  publisher = {American Physical Society},
  doi = {10.1103/PhysRevLett.124.210601},
  url = {https://link.aps.org/doi/10.1103/PhysRevLett.124.210601}
}

@article{Bibak2023,
  title = {Dissipative phase transitions in optomechanical systems},
  author = {Bibak, Fatemeh and Deli\ifmmode \acute{c}\else \'{c}\fi{}, Uro\ifmmode \check{s}\else \v{s}\fi{} and Aspelmeyer, Markus and Daki\ifmmode \acute{c}\else \'{c}\fi{}, Borivoje},
  journal = {Phys. Rev. A},
  volume = {107},
  issue = {5},
  pages = {053505},
  numpages = {8},
  year = {2023},
  month = {May},
  publisher = {American Physical Society},
  doi = {10.1103/PhysRevA.107.053505},
  url = {https://link.aps.org/doi/10.1103/PhysRevA.107.053505}
}

@article{Tomadin2012,
  title = {Reservoir engineering and dynamical phase transitions in optomechanical arrays},
  author = {Tomadin, A. and Diehl, S. and Lukin, M. D. and Rabl, P. and Zoller, P.},
  journal = {Phys. Rev. A},
  volume = {86},
  issue = {3},
  pages = {033821},
  numpages = {14},
  year = {2012},
  month = {Sep},
  publisher = {American Physical Society},
  doi = {10.1103/PhysRevA.86.033821},
  url = {https://link.aps.org/doi/10.1103/PhysRevA.86.033821}
}

@article{Ferreri2023,
  title = {Quantum field heat engine powered by phonon-photon interactions},
  author = {Ferreri, Alessandro and Macr\`{\i}, Vincenzo and Wilhelm, Frank K. and Nori, Franco and Bruschi, David Edward},
  journal = {Phys. Rev. Res.},
  volume = {5},
  issue = {4},
  pages = {043274},
  numpages = {9},
  year = {2023},
  month = {Dec},
  publisher = {American Physical Society},
  doi = {10.1103/PhysRevResearch.5.043274},
  url = {https://link.aps.org/doi/10.1103/PhysRevResearch.5.043274}
}

@article{frisk_kockum_ultrastrong_2019,
	title = {Ultrastrong coupling between light and matter},
	volume = {1},
	issn = {2522-5820},
	url = {http://www.nature.com/articles/s42254-018-0006-2},
	doi = {10.1038/s42254-018-0006-2},
	number = {1},
	urldate = {2021-11-21},
	journal = {Nat. Rev. Phys.},
	author = {Frisk Kockum, Anton and Miranowicz, Adam and De Liberato, Simone and Savasta, Salvatore and Nori, Franco},
	month = jan,
	year = {2019},
	pages = {19--40},
}

@article{Forn2019,
  title = {Ultrastrong coupling regimes of light-matter interaction},
  author = {Forn-D\'{\i}az, P. and Lamata, L. and Rico, E. and Kono, J. and Solano, E.},
  journal = {Rev. Mod. Phys.},
  volume = {91},
  issue = {2},
  pages = {025005},
  numpages = {48},
  year = {2019},
  month = {Jun},
  publisher = {American Physical Society},
  doi = {10.1103/RevModPhys.91.025005},
  url = {https://link.aps.org/doi/10.1103/RevModPhys.91.025005}
}

@article{zhou2025microscopic,
  title={Microscopic theory of heat transfer across a vacuum},
  author={Zhou, Yue-Hui and Liao, Jie-Qiao},
  journal={arXiv preprint arXiv:2508.02351},
  year={2025},
  doi={10.48550/arXiv.2508.02351}
}
\newpage

\appendix

\begin{center}
    \textbf{Supplementary Information for
``Explanation of the Observed Energy Exchange through the vacuum in Optomechanics"}
\end{center} 

\tableofcontents

\section{The system Hamiltonian}\label{app: SystemHamiltonian}
The system under consideration is a three-dimensional cavity-optomechanics consisting of two movable membrane located along the $x$-axis at $x_1$ and $x_2=x_1+d$. The confined field is expressed in terms of a massless scalar field $\phi(t,\mathbf{x})$ whose Lagrangian density is
\begin{align}\label{lagrangian:density}
\mathcal{L}(t,\mathbf{x})=\frac{1}{2}\partial_\mu\phi(t,\mathbf{x})\partial^\mu\phi(t,\mathbf{x}).
\end{align}
A central feature of cavity optomechanical systems is that the boundaries defining the electromagnetic field confinement are not fixed, but instead fluctuate around their equilibrium positions. Since the cavity resonance frequencies depend explicitly on the cavity length, any mechanical motion of the membrane directly affects the field dynamics. We denote the classical oscillation amplitudes of the two movable boundaries along the $x$ direction by $\delta x_1$ and $\delta x_2$, corresponding to equilibrium positions $x_1$ and $x_2$, respectively. In the regime of small displacements, $\delta x_1 \ll x_1$ and $\delta x_2 \ll x_2$, the boundary motion can be treated perturbatively. This approximation enables a systematic expansion of the field energy in the mechanical displacements and, following the procedure outlined in \cite{Ferreri2022a}, leads to the Hamiltonian of the coupled optomechanical system.

The first step of the procedure consists in solving the equation of motion for the classical scalar field under static boundary conditions imposed by the two cavity membrane. Specifically, we consider the Klein–Gordon equation,
\be
\partial_t^2\phi-\nabla^2\phi=0 
\ee
supplemented by Dirichlet boundary conditions at the fixed membrane positions. The field can then be expressed in terms of its normal modes through a Fourier decomposition compatible with these boundary conditions
\begin{align}\label{field:expression}
\phi(t,\mathbf{x})=&\sum_n^{\infty}\int d^2{\mathbf{k}}_\perp\left[\alpha_{\mathbf{k}}\,\phi_{\mathbf{k}}(t,\mathbf{x})+\alpha_{\mathbf{k}}^*\,\phi^*_{\mathbf{k}}(t,\mathbf{x})\right],
\end{align}
where each mode $\phi_{\mathbf{k}}(t,\mathbf{x})$ is described by
\begin{align}\label{field:modes}
\phi_{\mathbf{k}}(t,\mathbf{x})=&\frac{e^{i \mathbf{k}_\perp\cdot \mathbf{x}_\perp}}{2\pi\sqrt{\omega_{\mathbf{k}} d}}\sin\left[k_n(x-x_1)\right]e^{-i\omega_nt}.
\end{align}
In the expression above, we have defined the classical field amplitude $\alpha_{\mathbf{k}}$, the vector $\mathbf{x}\equiv (x,\mathbf{x}_\perp)= (x,y,z)$, and the wave vector $\mathbf{k}\equiv(k_n,\mathbf{k}_{\perp})=(\frac{n\pi}{d},k_y,k_z)$. The dispersion relation reads $\omega_n=c\sqrt{k_n^2+k_\perp^2}$, with $c$ is the velocity of light. By applying standard procedures, we can calculate the Hamiltonian density $\mathcal{H}(t,\mathbf{x})=\frac{1}{2}\left[(\Pi(t,\mathbf{x}))^2+(\nabla\phi(t,\mathbf{x}))^2\right]$, with canonical momentum $\Pi(t,\mathbf{x}):=-\partial_t\phi(t,\mathbf{x})$. This reads
%\begin{widetext}
\begin{align}\label{classical_Hamiltonian}
\mathcal{H}(t,&\mathbf{x})=\sum_{n\,m}^{\infty}\int \frac{d^2{\mathbf{k}}_\perp d^2\mathbf{k}'_\perp}{2(2\pi)^2 d}\left\{\frac{c^2k_n k_m}{\sqrt{\omega_{\mathbf{k}}\omega_{\mathbf{k}'}}}\left(\alpha_{\mathbf{k}}e^{i \mathbf{k}_{\perp}\cdot\mathbf{x}_\perp}+\alpha_{\mathbf{k}}^*e^{-i \mathbf{k}_{\perp}\cdot\mathbf{x}_\perp}\right)\left(\alpha_{\mathbf{k}'}e^{i \mathbf{k}'_\perp\cdot\mathbf{x}_\perp}+\alpha_{\mathbf{k}'}^*e^{-i \mathbf{k}'_\perp\cdot\mathbf{x}_\perp}\right)c_nc_m\right.\nonumber\\
&\left.-\left(\frac{c^2k_y k'_y+c^2k_z k'_z-\omega_{\mathbf{k}}\omega_{\mathbf{k}'}}{\sqrt{\omega_{\mathbf{k}}\omega_{\mathbf{k}'}}}\right)\left(\alpha_{\mathbf{k}}e^{i \mathbf{k}_{\perp}\cdot\mathbf{x}_\perp}-\alpha_{\mathbf{k}}^*e^{-i \mathbf{k}_{\perp}\cdot\mathbf{x}_\perp}\right)\left(\alpha_{\mathbf{k}'}e^{i \mathbf{k}'_\perp\cdot\mathbf{x}_\perp}-\alpha_{\mathbf{k}'}^*e^{-i \mathbf{k}'_\perp\cdot\mathbf{x}_\perp}\right)s_ns_m\right\}
\end{align}
%\end{widetext}
where $c_n =\cos\left[k_n(x-x_1)\right]$ and $s_n =\sin\left[k_n(x-x_1)\right]$. Once obtained the Hamiltonian density, the procedure \cite{Ferreri2022a} to quantize the optomechanical system in \eqref{classical_Hamiltonian} proceeds as follows:
\begin{enumerate}
    \item Induce a small fluctuation of the average position of the walls by replacing $x_i\rightarrow x_i+\delta x_i$ ($i=1,2$), and then expanding each wave vector $k_n$ in the Hamiltonian density with respect to the perturbation parameters $\epsilon_i=\delta x_i/d$ up to the desired order.
    \item Integrate the Hamiltonian density over the volume.
    \item Promote the field mode amplitudes $\alpha_{\mathbf{k}}$ and the small perturbations $\delta x_i$ to the quantum operators $\hat a_{\mathbf{k}}$ and $\delta x_i(\hat b_i+\hat b_i^\dag)$, respectively. 
\end{enumerate}
%We do not go through the first and the second steps into details, as they are lengthy calculations merely consisting of Maclaurin expansions and integrations over the spatial variables.In particular, the last step consists in imposing canonical bosonic commutation rules for the oscillation operators $[\hat b_i,\hat b_j^\dag]=\delta_{ij}$, as well as for the field modes $[\hat a_{\bold k},\hat a_{\bold k}^\dag]=(2\pi)^2(2\omega_{\bold k})\delta_{nm}\delta(\bold k_\perp-\bold{ k}'_\perp)$.
The first two steps follow from a systematic expansion of the relevant quantities in the mechanical displacements, combined with the integration over the spatial degrees of freedom. These standard but algebraically involved manipulations lead to the effective interaction terms of interest.
The final step consists in promoting the classical amplitudes to operators and imposing the canonical commutation relations. In particular, the mechanical modes satisfy the bosonic relations $[\hat b_i,\hat b_j^\dag]=\delta_{ij}$, as well as for the field modes $[\hat a_{\mathbf{n}},\hat a_{\mathbf{m}}^\dag]=\delta_{nm}\delta(\mathbf{k}_\perp-\mathbf{k}'_\perp)$ where we redefine $\hat{a}_{\mathbf{n}} \equiv \hat{a}_{\mathbf{k}}$ with $\mathbf{n}\equiv(k_n,\mathbf{k}_{\perp})$. The procedure ultimately leads to the Hamiltonian operator, which—upon expanding up to first order in the mechanical displacements—can be written as $\hat{H}=\hat H_0 +\hat H_{1}$ where
%\begin{widetext}
\begin{align}\label{quantum:hamiltonian:terms}
\hat{H}_0 =&\hbar\sum_{ n}^{\infty}\int d^2\mathbf{k}_\perp\omega_{ n}\,\bigg(\hat{a}_{\mathbf{n}}^\dag \hat{a}_{\mathbf{n}} +\frac{1}{2}\bigg)+\hbar\Omega_1\left(\hat{b}_1^\dag\hat{b}_1+\frac{1}{2}\right)+\hbar\Omega_2 \left(\hat{b}_2^\dag\hat{b}_2+\frac{1}{2}\right)\nonumber\\
\hat{H}_{I} =&\frac{\hbar c^2}{2}\sum_{nm}^{\infty}(-1)^{n+m}\int d^2\mathbf{k}_\perp\frac{k_n k_m}{\sqrt{\omega_{n}\,\omega_{m}}} (\hat{a}_{\mathbf{n}}^\dagger + \hat a_{\mathbf{n}})(\hat{a}_{\mathbf{m}}^\dagger + \hat a_{\mathbf{m}})\left(\epsilon_1(\hat{b}_{1}^\dagger + \hat b_{1})-\epsilon_2(\hat{b}_{2}^\dagger + \hat b_{2})\right).
%\hat{H}_{I} =&4\sum_{nm}(-1)^{n+m}\int d^2\bold k_\perp\frac{k_n k_m}{\sqrt{\omega_{ n}\,\omega_{ m}}} \hat X_{ n}\hat X_{ m}\left(\epsilon_1\hat Q_1-\epsilon_2\hat Q_2\right),
%\hat{H}_{2} =&4\sum_{nm}(-1)^{n+m}\int d^2\bold k_\perp\frac{k_n k_m}{\sqrt{\omega_{ n}\,\omega_{ m}}}\zeta_{ n  m}\hat X_{ n}\hat X_{ m}\left(\epsilon_1\hat Q_1-\epsilon_2\hat Q_2\right)^2,
\end{align}
%\end{widetext}
%with the usual definition of the canonical position operators  $\hat X_n=\frac{1}{2}(\hat a_n+\hat a_n^\dag)$ and $\hat Q_j=\frac{1}{2}(\hat b_j+\hat b_j^\dag)$.
%\begin{align}
%\zeta_{nm}=&2+k_{\perp}^2
%\frac{\omega_{ n}^2+\omega_{ m}^2}{(\omega_{ n}\omega_{ m})^2},
%\end{align}

\section{Effective Hamiltonian with the generalized James’ method}
\label{app: EffectiveH}

For interacting quantum systems with large detuning, an effective Hamiltonian can be derived using the generalized James' method~\cite{Shao2017}. The system Hamiltonian is first recast in the interaction picture, isolating the time dependence associated with the resonance conditions,
\begin{equation}
\label{APP:Heff0} 
\hat H_{\rm I}(t) = \sum_{k} \left[ \hat h_k e^{-i \nu_k t} + \hat h_k^\dag e^{ i \nu_k t} \right] \, .    
\end{equation}
Here,  the $\nu_k$ is a combination of the bare transitions frequencies which account for calculating the effective second-order contribution, 
\begin{equation}
\label{APP:second-oredr}
\hat H_{\rm {eff}}^{(2)} (t) = - \frac{1}{\hbar}\sum_{j,k}  \left[ \hat h_j \hat h_k^\dag \frac{e^{-i ( \nu_j -\nu_k) t}}{\nu_k}  - \hat h_j^\dag \hat h_k \frac{e^{ i (\nu_j - \nu_k) t}}{\nu_k} \right] \, .    
\end{equation}
The method relies on the rotating-wave approximation (RWA), whereby all nonzero frequency contributions can be neglected. One therefore identifies the frequencies $\nu_{(j,k)}$  such that only the terms in $\hat H_{\rm eff}^{(2)} (t)$ whose exponential factors sum to zero are retained.
Starting from \eqref{quantum:hamiltonian:terms}, and assuming the two phononic degrees of freedom to be resonant $\Omega=\Omega_1=\Omega_2$ (much less than any frequency mode of the cavity), we obtain the following James operators, 
\begin{align}\label{h_i}
  \hat{h}_1 & =  \hbar c^2  \sum_{nm}^{\infty} (-1)^{n+m} \int d^2\mathbf{k}_\perp \frac{k_n k_m}{\sqrt{\omega_{ n}\,\omega_{ m}}} \hat{a}_{\mathbf{n}} \hat{a}_{\mathbf{m}} (\epsilon_1 \hat{b}_1 - \epsilon_2 \hat{b}_2)  && \nu_1=-\omega_n - \omega_m- \Omega\nonumber ,\\
  \hat{h}_2 &=  \hbar c^2 \sum_{nm}^{\infty} (-1)^{n+m} \int d^2\mathbf{k}_\perp \frac{k_n k_m}{\sqrt{\omega_{ n}\,\omega_{ m}}} \hat{a}_{\mathbf{n}} \hat{a}_{\mathbf{m}} (\epsilon_1 \hat{b}^\dagger_1 - \epsilon_2 \hat{b}^\dagger_2)  && \nu_2=-\omega_n - \omega_m +  \Omega \nonumber ,\\
  \hat{h}_3 &=  \hbar c^2 \sum_{nm}^{\infty} (-1)^{n+m} \int d^2\mathbf{k}_\perp \frac{k_n k_m}{\sqrt{\omega_{ n}\,\omega_{ m}}} \hat{a}_{\mathbf{n}} \hat{a}^\dagger_{\mathbf{m}} (\epsilon_1 \hat{b}_1 - \epsilon_2 \hat{b}_2)  && \nu_3=-\omega_n + \omega_m- \Omega\nonumber, \\
   \hat{h}_4 &=  \hbar c^2 \sum_{nm}^{\infty} (-1)^{n+m} \int d^2\mathbf{k}_\perp \frac{k_n k_m}{\sqrt{\omega_{ n}\,\omega_{ m}}} \hat{a}^\dagger_{\mathbf{n}} \hat{a}_{\mathbf{m}} (\epsilon_1 \hat{b}_1 - \epsilon_2 \hat{b}_2)  && \nu_4= \omega_n - \omega_m- \Omega \;.
\end{align}

Since the structure of the effective Hamiltonian in Eq.~(\ref{APP:second-oredr}) involves several combinations of the James operators, it is necessary to consider different pairs of indices. Consequently, multiple resonance conditions may arise; when satisfied, they could lead to different effective Hamiltonians. In particular, we identify two subsets of $\hat{h}$ pair combinations. 
As first, we observe that the subset consisting of time-dependent pair combinations, namely $\{\hat{h}_1\hat{h}_2^{\dagger},\hat{h}_2\hat{h}_3^{\dagger},\hat{h}_2\hat{h}_4^{\dagger}\}$, cannot be reduced to commutators as the associated resonance conditions cannot be satisfied. Therefore, these terms remain rapidly oscillating and can be neglected within the RWA. Indeed, the resonance conditions would require relations of the form $\pm (\omega_n + \omega_m) \mp (\omega_k \mp \omega_j)= 2\Omega$. However, such conditions cannot be fulfilled since 
\begin{equation}
    \pm |\omega_n + \omega_m| \mp |\omega_k \mp \omega_j| > c\pi d^{-1}(\pm |n + m| \mp |k \mp j|) \gg 2 \Omega \;,
\end{equation}
implying that the corresponding oscillation frequencies are far detuned from $\Omega$. A second subset yields time-independent commutators when the corresponding resonance condition is satisfied, 
\begin{align}\label{h_i_time-independent_combination}
  [\hat{h}_1, \hat{h}_1^{\dagger}]  &  &&  [\hat{h}_2, \hat{h}_2^{\dagger}] && \omega_n + \omega_m =\omega_k + \omega_j \nonumber ,\\
  [\hat{h}_3,\hat{h}_3^{\dagger}]&  && [\hat{h}_4,\hat{h}_4^{\dagger}]  && \omega_n - \omega_m =\omega_k - \omega_j \nonumber ,\\
  [\hat{h}_1,\hat{h}_3^{\dagger}] & && [\hat{h}_1,\hat{h}_4^{\dagger}] && \omega_n + \omega_m =\omega_k - \omega_j, \nonumber  \\
   [\hat{h}_3,\hat{h}_4^{\dagger}] & && && \omega_n + \omega_m =-\omega_k + \omega_j,
\end{align}
leading to four distinct effective Hamiltonians,   
\begin{align}\label{h_i_time-independent_combination1}
  \hat{H}^{(2)}_{\rm eff_1}=& -\frac{1}{\hbar}\sum_{\alpha=1}^2  \comm{\hat{h}_{\alpha}}{\frac{1}{\nu_{\alpha}}\hat{h}_{\alpha}^\dagger}  &&  \omega_n + \omega_m =\omega_k + \omega_j \nonumber,\\
  \hat{H}^{(2)}_{\rm eff_2}=&-\frac{1}{\hbar}\sum_{\alpha=3}^4  \comm{\hat{h}_{\alpha}}{\frac{1}{\nu_{\alpha}}\hat{h}_{\alpha}^\dagger}  &&  \omega_n - \omega_m =\omega_k - \omega_j \nonumber ,\\
  \hat{H}^{(2)}_{\rm eff_3}=&-\frac{1}{\hbar}\sum_{\alpha=3}^4  \comm{\hat{h}_1}{\frac{1}{\nu_{\alpha}}\hat{h}_{\alpha}^\dagger}  && \omega_n + \omega_m =\omega_k - \omega_j, \nonumber  \\
   \hat{H}^{(2)}_{\rm eff_4}=&-\frac{1}{\hbar}[\hat{h}_3,\frac{1}{\nu_{4}} \hat{h}_4^{\dagger}]  &&  \omega_n + \omega_m =-\omega_k + \omega_j.
\end{align}
These resonance conditions are mutually exclusive; satisfying one necessarily excludes the others. For $(n,m)\neq (k,j)$, and using the bosonic commutation relations $[\hat b_i,\hat b_j^\dagger]=\delta_{ij}$, together with those for the field modes, $[\hat a_{\mathbf{n}},\hat a_{\mathbf{m}}^\dagger]=\delta_{nm}~\delta(\mathbf{k}_\perp-\mathbf{k}'_\perp)$, each resonance condition in~\eqref{h_i_time-independent_combination1} yields commutators that give rise to effective Hamiltonians describing purely photonic scattering processes, with no contribution from the mechanical degrees of freedom.

Consequently, coherent mechanical energy exchange can only be achieved by imposing $(n,m) = (k,j)$; in particular this requires $n = m$ for any value of the wave vector $\mathbf{k}_\perp$ (note that each discrete index involved is not in bold). Under this constraint, the interaction part of the system Hamiltonian in~\eqref{quantum:hamiltonian:terms} reduces to
\begin{equation}\label{quantum:hamiltonian:terms2}
\hat{H}_{I} =  \frac{\hbar c^2}{2}\sum_{n}^{\infty}\int d^2\mathbf{k}_\perp\frac{k_n^2}{\omega_n} \left[\hat a_{\mathbf{n}}^{\dagger 2} + \hat a_{\mathbf{n}}^2 + 2 \hat a_{\mathbf{n}}^\dagger \hat a_{\mathbf{n}} + \delta(\mathbf{k}_\perp - \mathbf{k}'_\perp) \right]\left[(\epsilon_1 \hat{b}_{1}^\dagger - \epsilon_2 \hat b_{2}^\dagger) + (\epsilon_1 \hat{b}_{1} - \epsilon_2 \hat b_{2}) \right].
%\hat{H}_{2} =&4\sum_{nm}(-1)^{n+m}\int d^2\bold k_\perp\frac{k_n k_m}{\sqrt{\omega_{ n}\,\omega_{ m}}}\zeta_{ n  m}\hat X_{ n}\hat X_{ m}\left(\epsilon_1\hat Q_1-\epsilon_2\hat Q_2\right)^2,
\end{equation}
Accordingly, the James operators and the associated resonant frequencies are defined as 
\begin{align}\label{h_i2}
  \hat{h}_1 & =  \frac{\hbar c^2}{2} \sum_{n}^{\infty} \int d^2\mathbf{k}_\perp \frac{k_n^2}{\omega_{n}} \hat{a}_{\mathbf{n}}^2 (\epsilon_1 \hat{b}_1 - \epsilon_2 \hat{b}_2)  && \nu_1=-2\omega_n - \Omega\nonumber ,\\
  \hat{h}_2 &=  \frac{ \hbar c^2}{2} \sum_{n}^{\infty} \int d^2\mathbf{k}_\perp \frac{k_n^2}{\omega_{n}} \hat{a}_{\mathbf{n}}^2 (\epsilon_1 \hat{b}^\dagger_1 - \epsilon_2 \hat{b}^\dagger_2)  && \nu_2=-2 \omega_n +  \Omega \nonumber ,\\
  \hat{h}_3 &=  \frac{\hbar c^2}{2} \sum_{n}^{\infty} \int d^2\mathbf{k}_\perp \frac{k_n^2}{\omega_{n}} [2 \hat a_{\mathbf{n}}^\dagger \hat a_{\mathbf{n}} + \delta(\mathbf{k}_\perp - \mathbf{k}'_\perp)]  (\epsilon_1 \hat{b}_1 - \epsilon_2 \hat{b}_2)  && \nu_3= - \Omega,  
\end{align}
%the last two resonance conditions in~\eqref{h_i_time-independent_combination} cannot be fulfilled and can therefore be neglected within the RWA. The first two conditions then coincide, yielding the effective Hamiltonian
%\begin{align}\label{h_i_time-independent_combination}
%\hat{H}^{(2)}_{\rm eff}=\sum_{\alpha=1}^4 \frac{1}{\omega_{\alpha}} \comm{\hat{h}_{\alpha}}{\hat{h}_{\alpha}^\dagger},
%\end{align}
while the corresponding effective Hamiltonian can then be derived as 
\begin{equation}
\hat{H}^{(2)}_{\rm eff}=-\frac{1}{\hbar}\sum_{\alpha=1}^3  \comm{\hat{h}_{\alpha}}{\frac{1}{\nu_{\alpha}} \hat{h}_{\alpha}^\dagger} =  \hat{H}_{\rm shift}^{(2)} + \hat{H}_{\rm JI}^{(2)}    
\end{equation}
where 
\begin{align}\label{shift_term}
\hat{H}_{\rm shift}^{(2)} =& \frac{\hbar c^4 (\epsilon_1^2 + \epsilon_2^2)}{\Omega} \sum_{n}^{\infty} \int d^2\mathbf{k}_\perp \frac{k_n^4}{\omega_{n}^2} \Big (2\hat a_{\mathbf{n}}^\dagger \hat a_{\mathbf{n}} + \frac{A_\perp}{(2\pi)^2} \Big ) + \frac{2 \hbar c^4 (\epsilon_1^2 + \epsilon_2^2)}{\Omega} \sum_{n}^{\infty} \Big ( \int d^2\mathbf{k}_\perp \frac{k_n^2}{\omega_{n}} \hat a_{\mathbf{n}}^\dagger \hat a_{\mathbf{n}} \Big )^2 \nonumber \\
+ &  \hbar c^4  \sum_{n}^{\infty} \int d^2\mathbf{k}_\perp \frac{k_n^4}{\omega_{n}^3} \Big (2\hat a_{\mathbf{n}}^\dagger \hat a_{\mathbf{n}} + \frac{A_\perp}{(2\pi)^2} \Big ) \Big (  \epsilon_1^2 (\hat{b}_1^\dagger \hat{b}_1 + \frac{1}{2}) + \epsilon_2^2 (\hat{b}_2^\dagger \hat{b}_2 + \frac{1}{2})\Big ),
\end{align}
is a combination of photonic and phononic frequency shift plus a nonlinear photonic Kerr term, and    
\begin{align}\label{Interaction_term}
\hat{H}_{\rm JI}^{(2)} =& - \hbar c^4 \epsilon_1 \epsilon_2 \sum_{n}^{\infty} \int d^2\mathbf{k}_\perp \frac{k_n^4}{\omega_{n}^3} \Big (2\hat a_{\mathbf{n}}^\dagger \hat a_{\mathbf{n}} + \frac{A_\perp}{(2\pi)^2} \Big ) \Big ( \hat{b}_1^\dagger \hat{b}_2  + \hat{b}_1  \hat{b}_2^\dagger \Big ),
\end{align}
is the effective membrane-membrane interaction part which is responsible for the phononic thermalization. Both $\hat{H}_{\rm shift}^{(2)}$ and $\hat{H}_{\rm JI}^{(2)}$ are derived under the frequency condition $\omega_n \gg \Omega$, which characterizes the standard optomechanical regime. In addition, their derivation involves the renormalization of the Dirac delta function \cite{peskin2018}, $(\delta(\mathbf{k}_\perp - \mathbf{k}'_\perp))^2=(A_\perp/4\pi^2)\delta(\mathbf{k}_\perp - \mathbf{k}'_\perp)$ to account for the finite size of the system. Here, $A_\perp$ denotes the area of the membrane surface perpendicular to the discrete photonic mode.    

\section{Strong phonon-phonon coupling through Casimir interaction}
\label{app: coupling}
The total effective Hamiltonian, $\hat{H}_0  + \hat{H}_{\rm shift}^{(2)} +\hat{H}_{\rm JI}^{(2)} $ 
%given in \eqref{quantum:hamiltonian:terms}, \eqref{shift_term}, and \eqref{Interaction_term}, respectively, 
can be diagonalized, thereby allowing the Hilbert space to be decomposed into sectors with fixed photon number. Consequently, the phonon-phonon coupling arises from two distinct contributions: (i) $g_{\rm th}$ due to the electromagnetic energy confined between the two membranes, as quantified by the photon number operator, and (ii) $g_{\rm c}$ due to the Casimir energy of the vacuum field. Although the system is maintained under high vacuum conditions (below $10^{-6}$ Torr, following Ref.~\cite{fong_phonon_2019}), the membrane are kept around room temperature. Consequently, thermally populated phonon modes can populate the intracavity photonic modes through radiative mechanism. At thermal equilibrium, the resulting photon occupations follow the Bose–Einstein distribution, $\langle \hat a_{\mathbf{n}}^\dagger \hat a_{\mathbf{n}} \rangle = (e^{\hbar\omega_n /k_{\rm B} T_{\rm eff}}-1)^{-1}$. Here, \(T_{\mathrm{eff}}\) is defined by equating the total power radiated by the two membranes to that emitted by an equivalent blackbody of area \(A_1+A_2\). The Stefan–Boltzmann law then gives $$
T_{\mathrm{eff}}
=
\left[
\frac{
A_1\mu_1\left(T_1^{\mathrm{bath}}\right)^4
+
A_2\mu_2\left(T_2^{\mathrm{bath}}\right)^4
}{
A_1+A_2
}
\right]^{1/4},
$$
where \(T_i^{\mathrm{bath}}\), \(A_i\), and \(\mu_i\) denote, respectively, the bath temperature, emitting area, and emissivity of membrane \(i=1,2\), with each membrane assumed to be in local thermal equilibrium with its reservoir.  As a result, the thermal contribution to the coupling is
\begin{align}\label{thermal_coupling_term}
g_{\rm th} =&  \frac{c^4 \epsilon_1 \epsilon_2}{2\pi^2}  \sum_{n}^{\infty} \int d^2\mathbf{k}_\perp \frac{k_n^4}{\omega_{n}^3} \frac{1}{e^{\hbar\omega_n /k_{\rm B} T_{\mathrm{eff}}}-1}.
\end{align}
Since $e^{-\hbar\omega_n /k_{\rm B} T_{\mathrm{eff}}}<1$, the Bose-Einstein factor can be expanded as a geometric series.  Introducing polar coordinates in the transverse plane $d^2\mathbf{k}_\perp=2\pi k dk$, and defining $q=(k^2 + k_n^2)^{1/2}$, \eqref{thermal_coupling_term} becomes
\begin{align}\label{thermal_coupling_term2}
g_{\rm th} =&  \frac{c \epsilon_1 \epsilon_2}{\pi} \sum_{m=1}^{\infty} \sum_{n=1}^{\infty} k_n^4 \int_{k_n}^{\infty} dq \frac{e^{-m q \lambda_T}}{q^2} \equiv  \frac{c \epsilon_1 \epsilon_2}{\pi} \sum_{m=1}^{\infty} \sum_{n=1}^{\infty} f_m(n), 
\end{align}
where $\lambda_T=\hbar c /(k_{\rm B} T_{\mathrm{eff}}) $ denotes the thermal wavelength. 
The sum over the discrete longitudinal modes can then be evaluated using the Poisson summation formula,
\begin{align}\label{the_sum_over_n}
\sum_{n=1}^{\infty} f_m(n) =& - \frac{1}{2} f_m(0) + \sum_{l=-\infty}^{\infty}  \int_{0}^{\infty} dx f_m(x) e^{2\pi i l x}, 
\end{align}
where $f_m(x)=(\pi x/d)^4 \int_{\pi x/d}^{\infty} dq e^{-m q \lambda_T }/q^2$.  Since $f_m(0)=0$, the $l=0$ term yields the bulk thermal contribution, 
\begin{align}\label{the_sum_over_n1}
g_{\rm th} =&  \frac{\epsilon_1 \epsilon_2 A_{\perp} d }{75 c^3} \Big(\frac{k_{\rm B} T_{\mathrm{eff}}}{\hbar}\Big)^4 = \frac{c \epsilon_1 \epsilon_2 A_{\perp} d }{75} \Big(\frac{1}{\lambda_T}\Big)^4 , 
\end{align}
whereas the nonzero Poisson modes ($l\neq0$) generate a finite-size thermal correction associated with the discreteness of the spectrum,
\begin{align}\label{the_sum_over_m}
\Delta g_{\rm th} \sim &  \frac{\epsilon_1 \epsilon_2 A_{\perp} d }{c^3} \Big(\frac{k_{\rm B} T_{\mathrm{eff}}}{\hbar}\Big)^4 \sum_{m=1}^{\infty} e^{-2m d /\lambda_T}. 
\end{align}
Hence, the finite-size thermal correction is exponentially suppressed in the high-temperature (or small-separation) regime $T_{\mathrm{eff}} d\ll1$, corresponding to $\lambda_T\gg d$. Therefore, only the bulk thermal term contributes significantly to $g_{\rm th}$. 

The Casimir energy of the vacuum electromagnetic field provides the second contribution to the phonon-phonon coupling strength,
%In this case, \eqref{Interaction_term} reduces to
%\begin{align}\label{Casimir_Interaction_term}
%\hat{H}_{\rm JI}^{(2)} =& -  2 \pi^2  A_\perp \hbar c^4 \epsilon^2 \sum_{n} \int d^2\bold k_\perp \frac{k_n^4}{\omega_{n}^3} \Big ( \hat{b}_1^\dagger \hat{b}_2  + \hat{b}_1  \hat{b}_2^\dagger \Big ),
%\end{align}
\begin{equation}
g_{\rm c}= c^4 \epsilon_1 \epsilon_2  A_\perp  \sum_{n}^{\infty} \int \frac{d^2\mathbf{k}_\perp}{(2\pi)^2} \frac{k_n^4}{\omega_{n}^3}. 
\end{equation}
Using $\omega_n=c\sqrt{k_n^2+k_\perp^2}$, the transverse-momentum integral can be evaluated by observing that
\begin{equation}
\lim_{\alpha \to 1} \partial^2_\alpha \int \frac{d^2\mathbf{k}_\perp}{(2\pi)^2} (\alpha k_n^2+k_\perp^2)^{1/2} = - \frac{c^3}{4} \int \frac{d^2\mathbf{k}_\perp}{(2\pi)^2} \frac{k_n^4}{\omega_n^3}, 
\end{equation}
where $\alpha$ is a dimensionless parameter. One is therefore led to evaluate
\begin{equation}
g_{\rm c}= - 4 c \epsilon_1 \epsilon_2  A_\perp \lim_{\alpha \to 1} \partial^2_\alpha \sum_n^{\infty} \int \frac{d^2\mathbf{k}_\perp}{(2\pi)^2} (\alpha k_n^2+k_\perp^2)^{1/2}, 
\end{equation}
which is directly related to the Casimir energy density confined between the two membranes \cite{milton2001}. Evaluating the corresponding mode sum yields
\begin{equation}\label{ideal_coupling}
%g_c=   \frac{\pi^6}{30} c A_\perp \epsilon^2 \frac{1}{ d^3}, 
g_{\rm c}=   \frac{\pi^2}{240} c A_\perp \epsilon_1 \epsilon_2 \frac{1}{ d^3}, 
\end{equation}
%\begin{equation}
%g_c=  A_\perp \frac{\pi^6}{30} \frac{c \epsilon^2}{ d^3} =  \frac{\pi^6 }{30} A_\perp c  (\delta x)^2 \frac{1}{ d^5}, 
%\end{equation}
where $\epsilon_{(1,2)}=\delta x_{(1,2)}/d$ denotes the perturbation parameter which is directly related to the zero-point-fluctuation amplitude of the mechanical mode $\delta x_{(1,2)}=(\hbar/2\Omega m_{(1,2),\rm{eff}})^{1/2}$, being $m_{(1,2),\rm{eff}}$ the effective mass of the two different membrane (note that $\Omega=\Omega_1=\Omega_2$). The ratio of the thermal and Casimir contributions scales as $g_{\rm th}/g_{\rm c} \sim (d/\lambda_{T})^4$.  Hence, the relative importance of the two terms is set by the dimensionless parameter $d/\lambda_{T}$: in the regime $\lambda_{T} \gg d$, the Casimir contribution dominates, whereas for $\lambda_{T} \ll d$, the thermal contribution becomes dominant. Using the parameter values reported in Ref.~\cite{fong_phonon_2019}, the membrane-membrane coupling is found to be dominated by the Casimir contribution, such that the thermal contribution $g_{\mathrm{th}}$ can be neglected.

The Casimir-mediated phonon-phonon coupling was derived in the idealized limit of perfectly reflecting parallel plates. For two membranes of different area and effective mass separated, by a distance $d$, the corresponding coupling strength from \eqref{ideal_coupling} now reads
\begin{equation}
 g_{\rm c} =  \frac{\pi^2}{240} c A_{\mathrm{ov}} \frac{\hbar}{2\Omega\sqrt{m_{1,\mathrm{eff}}m_{2,\mathrm{eff}}}} \frac{1}{d^5}, \label{gc_ideal}
\end{equation}
where $A_{\mathrm{ov}}=280\times280~\mu\rm{m}^2$ denotes the effective overlap area between the two membranes, the effective masses are $m_{1,\mathrm{eff}}=4.8 \times 10^{-12}~\rm{Kg}$ and  $m_{2,\mathrm{eff}}=3.5 \times 10^{-12}~\rm{Kg}$, while the membrane frequency is $\Omega/2\pi=191.6~\rm{KHz}$. The experimental setup of Ref.~\cite{fong_phonon_2019}, however, involves realistic multilayer dielectric with finite reflectivity, and metallic nanostructures rather than perfect mirrors. In this case, the Casimir interaction must be evaluated within Lifshitz theory by introducing an appropriate correction factor \cite{lifshitz1992theory,milton2001},  where material dispersion and finite reflectivity enter through frequency-dependent Fresnel coefficients. The resulting Casimir energy is reduced relative to the ideal case, leading to a renormalized coupling of the form
\begin{equation}
g(d,T) = \chi(d,T)\, g_{\rm c}(d),
\end{equation}
where $\chi(d,T)$ encodes multilayer optical response, thermal effects via Matsubara summation, and polarization-dependent screening. For the material parameters relevant to Ref.~\cite{fong_phonon_2019} (bulk Si$_3$N$_4$ coated with gold), one finds $\chi(d,T)\ll 1$ over the experimentally accessible distance range, typically corresponding to a suppression by one to two orders of magnitude relative to the ideal-metal limit, $10^{-2}\lesssim \chi(d)\lesssim 10^{-1}$, hence $g \sim (10^{-2}-10^{-1})g_{\rm c}$. In particular, over the experimentally accessible separation range, the correction factor is approximately well described by
\begin{equation}\label{eq:etafit}
\chi \simeq 0.045 \left(\frac{300~{\rm nm}}{d}\right)^{0.22}
\end{equation}
for $200~{\rm nm}\lesssim d\lesssim1200~{\rm nm}$. Combining Eqs.~(\ref{gc_ideal}) and (\ref{eq:etafit}) gives the
realistic Casimir-mediated phonon coupling
\begin{equation}
g \simeq \pi \left(\frac{300~{\rm nm}}{d} \right)^{5.22}~{\rm Hz}
\label{eq:gcrealfinal}
\end{equation}
which includes both the geometric correction due to the reduced overlap
area and the finite-reflectivity reduction predicted by Lifshitz theory.

\section{Equations of Motion and Thermalization for time-independent coupling}
\label{app: Dynamics}
Using the parameter values reported in Ref.~\cite{fong_phonon_2019}, the membrane-membrane coupling is found to be dominated by the Casimir contribution, allowing the thermal contribution $g_{\rm th}$ to be neglected. Consequently, the photonic degree of freedom can be traced out from the total effective Hamiltonian, including the noninteracting contribution, which now reads
\begin{align}\label{Final_hamiltonian}
\hat{H}_{\rm{eff}} =\hbar \Bigl(\Omega + \frac{\epsilon_1}{\epsilon_2} g  \Bigl) \Bigl(\hat{b}_1^\dag\hat{b}_1+\frac{1}{2}\Bigl)+ \hbar \Bigl(\Omega + \frac{\epsilon_2}{\epsilon_1} g \Bigl) \Bigl(\hat{b}_2^\dag\hat{b}_2+\frac{1}{2}\Bigl) - \hbar g \left(\hat{b}_{1}^\dagger \hat b_{2} + \hat{b}_{1} \hat b_{2}^\dagger \right),
\end{align}
where $g$ denotes the realistic Casimir-mediated phonon-phonon coupling given in \eqref{eq:gcrealfinal}. Each membrane is coupled to its own thermal bath, such that the dissipative dynamics is described by the Lindblad master equation
\begin{align}\label{dissipative_equation}
\dot{\hat{\rho}} =  \mathcal{L}\hat{\rho}= -\frac{i}{\hbar}[\hat{H}_{\rm{eff}}, \hat{\rho}] 
+ \sum_j^2 \frac{\gamma_j}{2} \Bigl[(n_j^{\mathrm{th}}+1) \mathcal{D}[\hat{b}_j]\hat{\rho} + n_j^{\mathrm{th}} \mathcal{D}[\hat{b}^{\dagger}_j]\hat{\rho}\Bigl],
\end{align}
where \(\mathcal{D}[\hat{\bullet}]\hat{\rho} = 2 \hat{\bullet} \hat{\rho} \hat{\bullet}^\dagger - \hat{\bullet}^\dagger \hat{\bullet} \hat{\rho} - \hat{\rho}\hat{\bullet}^\dagger \hat{\bullet} \) denotes the generic Lindblad dissipator, and the thermal phonon population is given by $n_j^{\mathrm{th}} = (e^{\hbar\Omega /k_{\rm B} T_j^{\rm bath}}-1)^{-1}$, where $T_j^{\rm bath}$ is the temperature of the $j$-th phononic bath.

In order to describe phononic thermalization as a signature of heat transfer mediated by quantum vacuum fluctuations, one can analyze the steady-state equations for the second-order correlation functions. By defining $B_j = \langle b_j^\dagger b_j \rangle$ with $j={(1,2)}$, and the coherence $C = \langle b_1^\dagger b_2 \rangle=R+iI$ (being $C \in \mathbb{C}$), the steady-state equations follow from the condition $\dot{\hat{O}}=\rm{Tr}[\hat{O}\dot{\hat{\rho}}]=\rm{Tr}[\hat{O}\mathcal{L}\hat{\rho}_{ss}]=0$, where $\hat{\rho}_{ss}$ denotes the steady-state density matrix corresponding to thermal equilibrium. Therefore, for $(B_1,B_2,I,R)^T_{ss}$, one obtains  
\begin{equation}\label{linear_system}
\begin{pmatrix}
-\gamma_1 & 0 & -2g & 0 \\
0 & -\gamma_2 & 2g & 0 \\
g & -g & -\Gamma & g\Delta\\
0 & 0 & -g\Delta & -\Gamma 
\end{pmatrix}
\begin{pmatrix}
B_1 \\
B_2 \\
I \\
R
\end{pmatrix}
= \begin{pmatrix}
-\gamma_1 n_1^{\mathrm{th}} \\
 -\gamma_2  n_2^{\mathrm{th}} \\
0 \\
0
\end{pmatrix}
\end{equation}
%\begin{equation}\label{linear_system}
%\begin{cases}
            %B_1 - n_1^{\mathrm{th}} + i  \dfrac{g}{\gamma_1}(C-C^*)=0 \\
            %\\
            %B_2 - n_2^{\mathrm{th}} - i \dfrac{ g}{\gamma_2}(C-C^*)=0\\
            %\\
            %C +   i \dfrac{2 g}{\gamma_1 + \gamma_2}(B_1-B_2)=0, 
        %\end{cases}
%\end{equation}
where $\Gamma= (\gamma_1 + \gamma_2)/2$ and $\Delta=(\epsilon_1^2-\epsilon_2^2)/\epsilon_1\epsilon_2= (m^2_{2,\mathrm{eff}}-m^2_{1,\mathrm{eff}})/m_{1,\mathrm{eff}}m_{2,\mathrm{eff}}$, being $m_{(1,2),\rm{eff}}$ the effective mass of the two different membrane (note that $\Omega=\Omega_1=\Omega_2$). The linear equation system in \eqref{linear_system}, when solved, yields 
\begin{equation}\label{Solution}
B_j= n_j^{\mathrm{th}} + \frac{\gamma_j (n_j^{\mathrm{th}} - n_k^{\mathrm{th}})}{(\gamma_k +  \gamma_j) \Bigl(1+\dfrac{\gamma_k \gamma_j}{4 g^2} + \Delta^2 \dfrac{ \gamma_k \gamma_j}{(\gamma_k + \gamma_j)^2} \Bigl)},
\end{equation}
with $j=(1,2)$ for $k=(2,1)$.  
\subsection{membrane-membrane thermalization}
Equation (\ref{Solution}) represents the thermal phonon occupation, $B_j={n'}^{\mathrm{th}}_j=(e^{\hbar\Omega /k_{\rm B} T_j^{'}}-1)^{-1}$, at an effective temperature $T_j^{'}$ which differs from the bath temperature $T_j^{'}\neq T_j^{\rm{bath}}$ due to phonon energy exchange between the membranes. Both phononic baths are maintained near room temperature, with one cooler than the other ($T_1^{\rm{bath}} < T_2^{\rm{bath}}$). Moreover, since the membrane frequency is of the order of $10^3$ Hz, \eqref{Solution} remains valid in the regime $\hbar \Omega/k_{\rm B} T\ll 1$, for which ${n}_{\mathrm{th}} \simeq k_{\rm B} T/\hbar \Omega$. In this limit, \eqref{Solution} reduces to
\begin{equation}\label{Thermalization}
T_j^{'}= T_j^{\rm{bath}} + \frac{\gamma_j (T_j^{\rm{bath}} - T_k^{\rm{bath}})}{(\gamma_k + \gamma_j) \Bigl(1+\dfrac{\gamma_k \gamma_j}{ \Tilde{g}^2} + \Delta^2 \dfrac{ \gamma_k \gamma_j}{(\gamma_k + \gamma_j)^2}\Bigl)},
\end{equation}
where $\Tilde{g}=2g$ and $j=(2,1)$ for $k=(1,2)$. Unlike the phenomenological model of Ref.~\cite{fong_phonon_2019}, our microscopic derivation naturally accounts for the asymmetry of the membrane effective masses through the additional term $ \Delta^2\gamma_k \gamma_j/(\gamma_k + \gamma_j)^2$, which originates from the fabrication-induced mass mismatch. Using the experimental parameters of Ref.~\cite{fong_phonon_2019}, we find $\Delta^2\gamma_k \gamma_j/(\gamma_k + \gamma_j)^2 \sim 10^{-2}$, so that this correction can be safely neglected, therefore  \eqref{Thermalization} reduces to the thermalization formula reported in Ref.~\cite{fong_phonon_2019}. In the weak-coupling regimes, $\Tilde{g} \ll \gamma_1,\gamma_2$, \eqref{Thermalization} reduces to $T_j^{'}= T_j^{\rm{bath}}$. Conversely, in the strong-coupling regime, $\Tilde{g} \gg \gamma_1,\gamma_2$, the two membranes thermalize to a common effective temperature,  $T_1^{'}= T_2^{'}= (\gamma_1 T_1^{\rm{bath}} + \gamma_2 T_2^{\rm{bath}})/(\gamma_1+ \gamma_2)$. 
%In \figref{fig: Comparison} we plot the comparison between the Casimir-mediated phonon coupling $g$ in \eqref{eq:gcrealfinal} and the scaling law  $g_{\text{Ref.~\cite{fong_phonon_2019}}} \propto d^{-4.91\pm 0.12}$. 
Remarkably, we find that $g_{_\text{\cite{fong_phonon_2019}}} \propto d^{-4.91\pm 0.12} \approx 2g $ (see \eqref{eq:gcrealfinal}), which accounts for the factor of two appearing in the thermalization formula \eqref{Thermalization}.
 
\section{Equations of Motion and Thermodynamics for time-dependent coupling }
\label{app: ThermoDynamics}
In this section, we discuss the validity of the adiabatic approximation used in the main text. To implement the thermodynamic cycle under consideration, we introduce a time-dependent phonon-phonon coupling $g(t)$, which acts as the external control parameter driving the system out of equilibrium. We then establish the relevant timescale hierarchy governing the modulation, derive the resulting non-equilibrium steady-state response, and evaluate the thermodynamic quantities associated with the driving cycle. 

The time dependence of the parametric coupling, $g(t)=g_0+A\sin(\theta t)$, originates from a controlled harmonic modulation of the physical distance between the two mechanical membranes. The modulation frequency $\theta$ is assumed to be much smaller than both the mechanical frequency $\Omega$ and the relaxation rates $\gamma_{(1,2)}$ associated with the thermal reservoirs, such that $\theta/(\gamma_{1}+\gamma_{2}) \ll 1$. Under this condition, the modulation is sufficiently slow to avoid non-adiabatic transitions and preserve the modal structure of the system, so that the mechanical configuration evolves quasi-statically. 

The validity of a time-dependent effective Hamiltonian $\hat{H}_{\rm{eff}}(t)$ relies on this separation of timescales between the internal relaxation dynamics and the external parametric driving. The fast timescale is set by the dissipation rates $\gamma_{(1,2)}$, which govern the exchange of energy between the nanomechanical modes and their respective thermal reservoirs, and is characterized by $\tau_{\rm int} \sim 1/\gamma_i$. The slow timescale is instead determined by the modulation frequency $\theta$, with driving period $T_{\rm ext}=2\pi/\theta$. The condition $\theta/(\gamma_{1}+\gamma_{2}) \ll 1$ therefore implies  $\tau_{\rm int} \ll T_{\rm ext}$. Physically, this relation ensures that the modes undergo many relaxation events during a single modulation cycle. As a result, the system continuously adapts to the slowly varying coupling $g(t)$, remaining close to a state of local thermodynamic equilibrium at each instant. The instantaneous normal modes of $\hat{H}_{\rm{eff}}(t)$ therefore remain well defined throughout the cycle, justifying the effective Hamiltonian description even in the presence of coupling modulation. Therefore, under the time-scale condition $\tau_{\rm int} \ll T_{\rm ext}$, \eqref{Final_hamiltonian} can be generalized to the time-dependent form
\begin{align}\label{Final_hamiltonian_time_dependent}
\hat{H}_{\rm{eff}}(t) =\hbar \Bigl(\Omega + \frac{\epsilon_1}{\epsilon_2} g(t)  \Bigl) \Bigl(\hat{b}_1^\dag\hat{b}_1+\frac{1}{2}\Bigl)+ \hbar \Bigl(\Omega + \frac{\epsilon_2}{\epsilon_1} g(t) \Bigl) \Bigl(\hat{b}_2^\dag\hat{b}_2+\frac{1}{2}\Bigl) - \hbar g(t) \left(\hat{b}_{1}^\dagger \hat b_{2} + \hat{b}_{1} \hat b_{2}^\dagger \right),
\end{align}
which provides the effective description of the coherent dynamics under slow modulation of the coupling strength. The dissipative dynamics is still described by the Lindblad master equation in \eqref{dissipative_equation}, and the time-dependent second-order correlation functions  are  defining by $B_j = \langle b_j^\dagger b_j \rangle$ with $j={(1,2)}$, and the coherence $C = \langle b_1^\dagger b_2 \rangle=R+iI$, being $C \in \mathbb{C}$. 

Defining the dynamic vector ${\mathbf{v}}=(B_1,B_2,I,R)^T_{t}$ 
%where $I=\rm{Im(C)}=Y$
, from the general equation $\langle \dot{\hat{O}}\rangle=\rm{Tr}[\hat{O}\dot{\hat{\rho}}]=\rm{Tr}[\hat{O}\mathcal{L}(t)\hat{\rho})]$ one obtains 
\begin{equation}\label{linear_system_time}
\begin{pmatrix}
\dot{B}_1 \\
\dot{B}_2 \\
\dot{I} \\
\dot{R}
\end{pmatrix}
=
\begin{pmatrix}
-\gamma_1 & 0 & -2 g(t) & 0 \\
0 & -\gamma_2 & 2 g(t)  & 0 \\
g(t) & -g(t) & -\Gamma & g(t) \Delta \\
0 & 0 & - g(t) \Delta & -\Gamma 
\end{pmatrix}
\begin{pmatrix}
B_1 \\
B_2 \\
I \\
R
\end{pmatrix}
+ 
\begin{pmatrix}
\gamma_1 n_1^{\mathrm{th}} \\
\gamma_2 n_2^{\mathrm{th}} \\
0 \\
0
\end{pmatrix}
\end{equation}
with $\Gamma=(\gamma_1+\gamma_2)/2$ and $\Delta=(\epsilon_1^2-\epsilon_2^2)/\epsilon_1\epsilon_2= (m^2_{2,\mathrm{eff}}-m^2_{1,\mathrm{eff}})/m_{1,\mathrm{eff}}m_{2,\mathrm{eff}}$, being $m_{(1,2),\rm{eff}}$ the effective mass of the two different membrane (note that $\Omega=\Omega_1=\Omega_2$). 
%From the last equation, it is evident that the real part $R$ of the coherence is completely decoupled from the rest of the dynamical variables and undergoes a passive exponential decay toward zero $R(t)=R(0) e^{-\Gamma t}$. Consequently, in the asymptotic long-time steady state, this component vanishes identically. This decoupling allows us to reduce the effective state space of the system to the three-dimensional vector ${\bold v}=(B_1,B_2,I)^T_{t}$, whose non-equilibrium dynamics are dictated by the equations system
%\begin{equation}\label{linear_system_2}
%\begin{pmatrix}
%\dot{B}_1 \\
%\dot{B}_2 \\
%\dot{I}
%\end{pmatrix}
%=
%\begin{pmatrix}
%-\gamma_1 & 0 & -2 g(t)  \\
%0 & -\gamma_2 & 2 g(t) \\
%g(t) & -g(t) & -\Gamma  
%\end{pmatrix}
%\begin{pmatrix}
%B_1 \\
%B_2 \\
%I 
%\end{pmatrix}
%+ 
%\begin{pmatrix}
%\gamma_1 n_1^{\mathrm{th}} \\
%\gamma_2 n_2^{\mathrm{th}} \\
%0 
%\end{pmatrix}
%\end{equation}
%or in vector form $\dot{\bold v}(t) = \bold{M}(t)\bold v (t) + \bold a$. 
To solve the time-dependent problem \eqref{linear_system_time}, we employ the method of multiple scales by rescaling time with the total dissipation rate $\Gamma$ and introducing the small adiabatic parameter $\zeta =\theta/\Gamma \ll 1$. This naturally defines two independent timescales: (i) the slow timescale $\tau = \zeta \Gamma t = \theta t $, which accounts for the gradual evolution induced by the modulation of the coupling, and (ii) the fast timescale $\tau/\zeta = \Gamma t$, which describes the rapid relaxation toward the instantaneous steady state. As a consequence, the time derivative expands as: $d/dt = \Gamma (\partial_{(\tau/\zeta)} + \zeta \partial_{\tau})$. Therefore, the state vector ${\mathbf{v}}_{\tau}$ can be expanded in powers of $\zeta$, namely,  ${\mathbf{v}}(\tau) = {\mathbf{v}}^{(0)}(\tau) + \zeta {\mathbf{v}}^{(1)}(\tau) + \mathcal{O}(\zeta^2)$. This ansatz allows to systematically isolate the dynamical contributions order by order. Indeed, substituting the expanded ${\mathbf{v}}$ vector in the scaled system equation \eqref{linear_system_time} and neglecting the terms $\sim \mathcal{O}(\zeta^2)$, in vector form one obtains 
\begin{equation}\label{linear_system_time_vector}
\dot{\mathbf{v}} = \Gamma(\partial_{(\tau/\zeta)} + \zeta \partial_{\tau}){\mathbf{v}}^{(0)} + \zeta  \Gamma \partial_{(\tau/\zeta)} {\mathbf{v}}^{(1)} = \mathbf{M}(\tau) ({\mathbf{v}}^{(0)} + \zeta {\mathbf{v}}^{(1)}) + \mathbf{a}.  
\end{equation}
%$\dot{\bold v}(\tau) = \Gamma \partial_{\tau_0} \bold{v} + \zeta  \Gamma \partial_{\tau_1} \bold{v} = \bold{M}(\tau)\bold{v}(\tau) + \bold{a}$
When one looks at the long-time behavior of the system, the fast transients completely die out. In the multi-scale framework, this means: $\partial_{(\tau/\zeta)}{\mathbf{v}}^{(0)}=0$ at the zeroth-order, as well as, at the first-order $\partial_{(\tau/\zeta)}{\mathbf{v}}^{(1)}=0$. The derivative does not vanish instantaneously at $t=0$. It vanishes because multi-scale perturbation theory specifically isolates the long-time physics, wherein the environment has already completed its  action on the system—bringing it to its local steady state—leaving only the slow non-adiabatic delay induced by $\theta$. Therefore, we end up with the following zeroth- and first-order correction vector equations:
\begin{equation}\label{vector_equation_system}
    \begin{cases}
       {\mathbf{v}}^{(0)}=-{[\mathbf{M}(\tau)]}^{-1}{\mathbf{a}} \\
        {\mathbf{v}}^{(1)}=\Gamma{[\mathbf{M}(\tau)]}^{-1}\partial_{\tau}{\mathbf{v}}^{(0)}\;.
    \end{cases}
\end{equation}
%(i) for the zeroth-order ${\bold v}^{(0)}=-\mathbf{M}^{-1}(t){\bold a}$, and (ii) for the first-order correction ${\bold v}^{(1)}=\Gamma\mathbf{M}^{-1}(t)\partial_{\tau}{\bold v}^{(0)}$. 
%Now that we have defined the multiscale formal framework using the slow time variable $\tau= \theta t$ and the tracking limit in which the fast derivatives vanish $\partial_{(\tau/\zeta)} \to 0  $, we can determine the tracking solution order by order.
\subsection{Zeroth-order solution}\label{ZOSolution}
The first vector equation of the system of \eqref{vector_equation_system} yields the instantaneous zeroth-order solutions: 
(i) the imaginary part of the coherence (heat current flow), 
\begin{equation}\label{I-Solution0}
I^{(0)}(\tau)= \frac{\gamma_1 \gamma_2(n_1^{\mathrm{th}}-n_2^{\mathrm{th}})}{2g(\tau)(\gamma_1 + \gamma_2) \bigg( 1+\dfrac{\gamma_1 \gamma_2}{4 g^2(\tau)} + \Delta^2 \dfrac{ \gamma_1 \gamma_2}{(\gamma_1 + \gamma_2)^2}\bigg)},
\end{equation}
(ii) the real part of the coherence 
\begin{equation}\label{R-Solution0}
R^{(0)}(\tau) = - \frac{g(\tau) \Delta}{\Gamma} I^{(0)}(\tau), 
\end{equation}
(iii) and the thermal population  $(j=1,2)$, 
\begin{equation}\label{B-Solution0}
B_j^{(0)}(\tau)= n_j^{\mathrm{th}} - \frac{2 g(\tau)}{\gamma_j} I^{(0)}(\tau).
\end{equation}
The zeroth-order solution (\eqref{B-Solution0} with \eqref{I-Solution0}) is, of course, the time-dependent version of \eqref{Solution}.
The zeroth-order solution represents a perfectly reversible, quasi-static configuration. The internal state tracks the instantaneous value of $g(\tau)$ without delay, implying that any energy absorbed during the compression phase is symmetrically returned during expansion.
\subsection{First-order correction}\label{first-Order_Solution}
The finite velocity of the external parameter $\dot{g}(\tau) = A\theta \cos(\tau) \neq 0$, driven by the non-zero but bounded modulation frequency $\theta \neq 0$, breaks the instantaneous equilibrium balance at the next order. Here, the time derivative of the zeroth-order solution $\partial_{\tau}{\mathbf{v}}^{(0)}(\tau) = \dot{\mathbf{v}}^{(0)}(\tau) = \dot{g}(\tau) \partial_g {\mathbf{v}}^{(0)}(\tau)$ acts as an effective non-equilibrium source term for the first-order correction $\mathcal{O}(\zeta)$. Consequently, the dynamical system reduces to the second equation of \eqref{vector_equation_system}, which once solved yields the first-order kinematic delay:
(i) the the first-order correction to the heat current flow, 
\begin{equation}\label{heat flow_correction}
I^{(1)}(\tau) = -\frac{\Gamma \dot{I}^{(0)}(\tau) + g(\tau) \Delta \dot{R}^{(0)}(\tau) + g(\tau) \Gamma \bigg( \dfrac{1}{\gamma_1} \dot{B}_1^{(0)}(\tau) - \dfrac{1}{\gamma_2} \dot{B}_2^{(0)}(\tau) \bigg)}{\dfrac{4 g^2(\tau) \Gamma}{\gamma_1 \gamma_2} \bigg( 1+\dfrac{\gamma_1 \gamma_2}{4 g^2(\tau)} + \Delta^2 \dfrac{ \gamma_1 \gamma_2}{(\gamma_1 + \gamma_2)^2}\bigg)} ,
\end{equation}
(ii) the first-order correction to the real part of the coherence 
\begin{equation}\label{R-Solution1}
R^{(1)}(\tau) =  -  \dot{R}^{(0)}(\tau) - \frac{g(\tau) \Delta}{\Gamma} I^{(1)}(\tau), 
\end{equation}
(iii) and the first-order correction to the thermal phononic populations  $(j=1,2)$, 
\begin{equation}\label{population_correction}
B_j^{(1)}(\tau)= - \frac{\Gamma}{\gamma_j}  \dot{B}_j^{(0)}(\tau)  - \frac{2 g(\tau)}{\gamma_j}  I^{(1)}(\tau).
\end{equation}
%Evaluating the derivative explicitly leads to the exact analytical expression for the kinematic lag:
%\begin{equation}
%\label{eq:current_first_order}
%I^{(1)}(t) = \theta \cdot \frac{8 \Gamma \gamma_1 \gamma_2 g(t) A \cos(\theta t) (n_1^{\text{th}} - n_2^{\text{th}})}{\left[\gamma_1 \gamma_2 + 4g^2(t)\right]^3}.
%\end{equation}
The latter equation demonstrate that the first-order population deviation is generated by \emph{two distinct physical contributions: the non-equilibrium delay} arising from tracking the instantaneous steady state $  \dot{B}_j^{(0)}(\tau)$, \emph{and the additional work or heat injected} by the out-of-phase current component $2 g(\tau) I^{(1)}(\tau)$.  Crucially, $I^{(1)}(\tau) \propto \cos(\tau)$ is not in phase with respect to the primary driving parameter $g(\tau) \propto \sin(\tau)$. This out-of-phase component explicitly breaks the time-reversal symmetry of the steady-state, mapping an open hysteresis loop in the state space that permits net heat-to-work conversion.

\subsection{Thermodynamics first-law Analysis and Closed-Form Cyclic Work Extraction}
\label{sec:first_law}
To evaluate the energetic performance of the nanomechanical heat engine, we consider the thermodynamic framework in open quantum systems theory. The instantaneous total energy of the system is given by the expectation value of the effective Hamiltonian, $E(\tau) = \text{Tr}[H(\tau)\rho(\tau)]$. Taking the total time derivative yields the quantum formulation of the First Law of Thermodynamics:
\begin{equation}\label{energy}
\dot{E}(\tau) = \text{Tr}\left[ \partial_{\tau} \hat{H}_{\rm{eff}}(\tau) \rho(\tau) \right] + \text{Tr}\left[\hat{H}_{\rm{eff}}(\tau) \dot{\rho}(\tau)\right] \equiv \dot{W}(\tau) + \sum_{j=1,2} \dot{Q}_j(\tau),
\end{equation}
where $\dot{W}(\tau)$ represents the mechanical power exchanged with the external control parameter, and $\dot{Q}_j(\tau)$ denotes the heat current flowing into the system from the $j$-th thermal reservoir. Using the explicit form of the time-dependent effective Hamiltonian in \eqref{Final_hamiltonian_time_dependent} one obtains:
\begin{equation}\label{power}
 \dot W(\tau) = \hbar\theta \dot{g}(\tau)  \left[ \frac{\epsilon_1}{\epsilon_2}\left(B_1(\tau)+\frac{1}{2}\right)  +  \frac{\epsilon_2}{\epsilon_1}\left(B_2(\tau)+\frac{1}{2}\right) -2R(\tau)  \right],
\end{equation}
where  $B_j(\tau) = B_j^{(0)}(\tau) + \zeta B_j^{(1)}(\tau) +  \mathcal O(\zeta^2)$ and $R(\tau) = R^{(0)}(\tau)  + \zeta R^{(1)}(\tau) + \mathcal O(\zeta^2)$ are the zeroth- and first-order solution of the thermal population and the real part of the coherence (see Sec.~\ref{ZOSolution}\textcolor{blue}{,2}), respectively. Expressed in terms of the general solution, \eqref{power} splits into a zeroth- and first-order mechanical power correction as follows
\begin{align}\label{power_separate}
 \dot W(\tau) =&\hbar\theta \dot{g}(\tau)  \bigg[  \frac{\epsilon_1}{\epsilon_2}B_1^{(0)}(\tau)  +  \frac{\epsilon_2}{\epsilon_1}B_2^{(0)}(\tau) -2R^{(0)}(\tau) +\frac{\Delta}{2} + \frac{\epsilon_2}{\epsilon_1}\bigg] \nonumber\\
 +& \hbar\theta \dot{g}(\tau) \zeta  \left[\frac{\epsilon_1}{\epsilon_2}B_1^{(1)}(\tau) +  \frac{\epsilon_2}{\epsilon_1}B_2^{(1)}(\tau) -2R^{(1)}(\tau)\right]+ \mathcal{O}(\zeta^2) \nonumber\\
 =& {\dot W}^{(0)}(\tau) + \zeta {\dot W}^{(1)}(\tau) + \mathcal{O}(\zeta^2).    
\end{align}

Neglecting the terms of the order $\sim \mathcal{O}(\zeta^2)$, the total work performed over one modulation period $T=2\pi/\theta$ is given by 
\begin{equation}\label{power_integral}
 W_{\rm cyc}=\int_0^{2\pi} d\tau \dot W(\tau)  = \int_0^{2\pi} d\tau {\dot W}^{(0)}(\tau) + \zeta \int_0^{2\pi} d\tau {\dot W}^{(1)}(\tau).
\end{equation}
The zeroth-order of the performed work is a single-valued function of the instantaneous coupling $g(\tau)$, therefore it does not give any contributions, 
\begin{equation}
W_{\rm cyc}^{(0)} =  \hbar  \theta \int_0^{2\pi} d\tau\, \dot{g}(\tau) \mathcal{W}(\tau)  = \hbar \theta  \oint dg \mathcal{W}(g)= 0,
\end{equation}
while the first non-vanishing cyclic work is generated by the first-order adiabatic delay of the system state with respect to the external modulation $g(\tau)=g_0 + A \sin{\tau}$. To compute the first non-vanishing cyclic work, we first rewrite the first-order correction ${\dot W}^{(1)}(\tau)$ in terms of the zeroth-order coherence solution \eqref{I-Solution0} by using \eqref{R-Solution0}-(\textcolor{blue}{E9}), 
%\begin{equation}\label{power_correction}
 %\dot W^{(1)}(\tau) = 2\hbar\theta \dot{g}(\tau)  \left[ \alpha \dot{g}(\tau) I^{(0)}(\tau)+ \alpha g(\tau) \partial_{\tau} I^{(0)}(\tau)  - \beta g(\tau)  I^{(1)}(\tau)  \right],
%\end{equation}
\begin{equation}\label{power_correction}
\zeta \dot W^{(1)}(\tau) = 2\hbar\theta^2 \dot{g}(\tau) \frac{d}{d\tau}\left[ \alpha \Phi(\tau) - \frac{\beta}{n_1^{\mathrm{th}}-n_2^{\mathrm{th}}}\bigg(k-\frac{1}{2g^2(\tau)}\bigg)\Phi^2(\tau) \right],
%=2\hbar\theta^2 \dot{g}(\tau) \frac{d}{d\tau} \mathcal{S}(\tau),
\end{equation}
where
\begin{align}
\Phi(\tau)=&g(\tau) I^{(0)}(\tau) \qquad \qquad \;\;\;  \alpha=\frac{\epsilon_1}{\epsilon_2 \gamma_1^2} - \frac{\epsilon_2}{\epsilon_1 \gamma_2^2} - \frac{\Delta}{\Gamma^2} \nonumber\\
\beta=&\frac{\epsilon_1}{\epsilon_2 \gamma_1} - \frac{\epsilon_2}{\epsilon_1 \gamma_2} - \frac{\Delta}{\Gamma} \qquad k=\frac{1}{\gamma_1^2 }+ \frac{1}{\gamma_2^2 } + \frac{\Delta^2}{2\Gamma^2}\nonumber. 
\end{align}
%where $\Phi(\tau)=g(\tau) I^{(0)}(\tau)$, $\alpha=\epsilon_1 /\epsilon_2 \gamma_1^2-\epsilon_2 /\epsilon_1 \gamma_2^2 - \Delta/\Gamma^2$, $\beta=\epsilon_1/\epsilon_2 \gamma_1-\epsilon_2 /\epsilon_1 \gamma_2 - \Delta/\Gamma$, and $k=1 /\gamma_1^2 + 1 /\gamma_2^2 + \Delta^2/2\Gamma^2$. 
Following the sign convention, the net work extracted from the system can be obtained integrating \eqref{power_correction} with a minus sign over a full period, 
\begin{equation}\label{final_formula}
W_{\rm cyc}^{\rm ext}=-\zeta \int_0^{2\pi} d\tau\,\dot W^{(1)}(\tau)= - 4\pi\hbar\frac{\theta}{\Gamma} (n_1^{\mathrm{th}}-n_2^{\mathrm{th}})s^2  \bigg[\bigg( \alpha  - \frac{\beta}{2\Gamma} \left(1+4ks\right) \bigg) X -  \frac{s \beta}{2\Gamma} \left(1+2ks\right)\frac{d X}{ds} \Bigg],
\end{equation}
where
%\begin{equation}
 %\mathcal{F}(g_0,A,\gamma_1,\gamma_2,\Delta) = s^4  \bigg[\bigg( 2 \alpha- \frac{\beta}{\Gamma} \left(1+4ks^2\right) \bigg) X -s \frac{\beta}{\Gamma} \left(\frac{1}{2}+ks^2\right)\frac{d X}{ds} \Bigg],
%\end{equation}
%with 
\begin{align}
s =&\frac{\gamma_1\gamma_2}{4+\dfrac{\Delta^2\gamma_1\gamma_2}{\Gamma^2}} \qquad \qquad \;\;\; \; u=\sqrt{(g_0^2-A^2-s)^2+4g_0^2s} \nonumber \\ 
r=& \sqrt{\frac{u + g_0^2 - A^2 - s}{2}} \qquad  X=\frac{r^2 - g_0^2}{r u}\nonumber. 
\end{align}

Interestingly, from \eqref{final_formula} we notice that 
\begin{equation}
 W_{\rm cyc}^{\rm ext} \propto -4\pi\hbar  \zeta (n_1^{\mathrm{th}}-n_2^{\mathrm{th}}) (\gamma_2^2-\gamma_1^2)\approx  -4\pi
\dfrac{k_{\rm B}}{\Omega}\zeta (T_1^{\rm{bath}}-T_2^{\rm{bath}}) (\gamma_2^2-\gamma_1^2),   
\end{equation}
%$W_{\rm cyc}^{\rm ext} \propto -4\pi\hbar  \zeta (n_1^{\mathrm{th}}-n_2^{\mathrm{th}}) (\gamma_2^2-\gamma_1^2)\approx  -4\pi\dfrac{k_B}{\Omega}\zeta (T_1^{\rm{Bath}}-T_2^{\rm{Bath}}) (\gamma_2^2-\gamma_1^2)$ 
in the regime $\hbar \Omega/k_{\rm B} T\ll 1$. Therefore the net work extracted over a full period, is proportional to $\zeta=\theta/\Gamma$, as expected for the first adiabatic correction to the mechanical power written in the slow variable $\tau$, and it depends on the temperature variation and on the difference between the heat losses through $\alpha$ and $\beta$. In particular, in order to get a net work extracted form the system, the sing minus has to be preserved, thus, one needs $(T_1^{\rm{bath}}-T_2^{\rm{bath}}) (\gamma_2^2-\gamma_1^2) >0 $. Assuming a negative external thermal gradient ($T_1^{\rm{bath}} < T_2^{\rm{bath}}$), the system functions as a quantum engine if $\gamma_2<\gamma_1$. This is not exactly the experimental case  studied in Ref.~\cite{fong_phonon_2019}, in which $\gamma_2>\gamma_1$. Nevertheless, the system used in Ref.~\cite{fong_phonon_2019} to study phonon heat transfer across quantum electromagnetic vacuum field remains a suitable experimental platform to test net work  extraction provided that the values of the dissipation rates can be inverted $\gamma_1>\gamma_2$. 

Having determined the work over a driving cycle, we now have one the ingredients required to evaluate the thermal efficiency. The other one we need in order to calculate thermal efficiency is the heat exchanged by the two membrane.  We have assumed from  Ref.~\cite{fong_phonon_2019} an  external thermal gradient ($T_1^{\rm{bath}} < T_2^{\rm{bath}}$), thus by denoting $Q_{\rm h}=Q_2>0$ the heat absorbed from the hot reservoir and using the net-work extracted  $W_{\rm cyc}^{\rm ext}$ over a cycle, the efficiency is given by $\eta=W_{\rm cyc}^{\rm ext}/Q_2$.  From \eqref{energy} one obtains
\begin{equation}\label{derivative_heat}
 \dot{Q}_j(\tau) = \hbar \gamma_j \left[  \bigg(\Omega + \frac{\epsilon_j}{\epsilon_k} g(\tau) \bigg) \bigg(n_j^{\mathrm{th}} - B_j(\tau) \bigg)+  g(\tau) R(\tau)  \right],
\end{equation}
where $k=(2,1)$ for $ j=(1,2)$. Using the zeroth- and first-order solution for the thermal population and the real part of the coherence (see \secref{ZOSolution}\textcolor{blue}{,2}), \eqref{derivative_heat} can be written in terms of  zeroth- and first-order mechanical heat correction as  $ \dot{Q}(\tau) =  {\dot Q}^{(0)}(\tau) + \zeta {\dot Q}^{(1)}(\tau) + \mathcal{O}(\zeta^2)$. The zeroth- and first-order solution mechanical heat correction can be express in terms of the only $I^{(0)}(\tau)$ (see \eqref{I-Solution0}) as follows 
\begin{align}\label{heat_separate}
{\dot Q}_j^{(0)}(\tau) =& 2 (-1)^{j+1} \hbar g(\tau) I^{(0)}(\tau) \bigg[\Omega + \bigg(\frac{\epsilon_j}{\epsilon_k}  + (-1)^{j} \frac{\gamma_j \Delta}{2 \Gamma} \bigg) g(\tau) \bigg], \nonumber\\
{\dot Q}_j^{(1)}(\tau)=& 2 (-1)^{j} \hbar \frac{\gamma_1 \gamma_2}{4 g^2(\tau) D} \dot{g}(\tau) I^{(0)}(\tau) \bigg[\frac{2 \Gamma}{\gamma_j} \bigg(\Omega + \frac{\epsilon_j}{\epsilon_k} g(\tau)\bigg) + (-1)^{j} \frac{\gamma_j \Delta}{\Gamma}g(\tau)\nonumber\\ 
+& (-1)^{j+1} \bigg[\bigg(\Omega + \frac{\epsilon_j}{\epsilon_k} g(\tau)\bigg) + (-1)^{j} \frac{\gamma_j \Delta}{2\Gamma}g(\tau)\bigg] \Lambda,    
\end{align}
with
\begin{align}
D =1+\frac{\gamma_1 \gamma_2}{4 g^2(\tau)} + \Delta^2 \frac{ \gamma_1 \gamma_2}{4\Gamma^2}, \qquad \Lambda=1-\frac{\gamma_1 \gamma_2}{2 g^2(\tau) D} + \frac{k \gamma_1 \gamma_2}{D}\nonumber. 
\end{align}
Note that, the zeroth-order heat current satisfies $\dot{Q}^{(0)}_1+\dot{Q}^{(0)}_2=0$, which represents the instantaneous stationary heat transport between the two thermal baths. 
%\begin{figure*}[t]
   % \centering
    %\includegraphics[width=0.6\textwidth]{efficiency.pdf}
    %\caption{Here, one plots the efficiency as function of the losses ratio $\gamma_1/\gamma_2$, and for different value of the adiabatic parameter $\zeta=\theta/\Gamma$. Efficiency increases when the loss asymmetry does, as well as increasing $\zeta$,  which makes faster the coupling-modulation $g(\theta t)=g_0+A\sin(\theta t)$ with $g_0= 1.57~\rm{Hz}$, and $A= 1.49~\rm{Hz}$. The other system parameters are taken from  Ref.~\cite{fong_phonon_2019}, $\Delta=0.31,~\Omega/2\pi= 191.6 \times 10^{3}~\rm{Hz},~T_{1}^{\rm{Bath}}=287~K, \rm{and}~T_{2}^{\rm{Bath}}=312.5~K$.}
    %\label{fig: efficiency}
%\end{figure*}
The heat current flowing into the system from the $j$-th thermal reservoir is obtained by integrating over one modulation period $T=2\pi/\theta$  
\begin{equation}\label{heat_integral}
 Q_j=\int_0^{2\pi} d\tau \dot{Q}_j(\tau)  = \frac{1}{\theta}\int_0^{2\pi} d\tau {\dot Q}_j^{(0)}(\tau) + \frac{1}{\Gamma} \int_0^{2\pi} d\tau {\dot Q}_j^{(1)}(\tau).
\end{equation}
While the zeroth-order heat current is easily integrable yielding,
\begin{equation}\label{heat_formula}
Q_j^{(0)}= (-1)^{j+1} \frac{2\pi \hbar}{\theta \Gamma}(n_1^{\mathrm{th}}-n_2^{\mathrm{th}})s \bigg[2 \Omega \bigg( 1  - \frac{ \sqrt{s}r_-}{u} \bigg) + \mu \bigg( g_0 - \frac{s r_+}{u} \Bigg)\Bigg],
\end{equation}
with 
\begin{align}
s =&\frac{\gamma_1\gamma_2}{4+\dfrac{\Delta^2\gamma_1\gamma_2}{\Gamma^2}} \qquad \qquad \qquad \;  u=\sqrt{(g_0^2-A^2-s)^2+4g_0^2s}\nonumber\\
r_{\pm}=&\sqrt{\frac{u \pm (g_0^2 - A^2 - s)}{2}}\; \; \qquad\mu=2\frac{\epsilon_1}{\epsilon_2} -  \frac{\gamma_1 \Delta}{\Gamma}= 2\frac{\epsilon_2}{\epsilon_1}+\frac{\gamma_2 \Delta}{\Gamma}\nonumber.  
\end{align}
Surprisingly, the integrated first-order does not give any contributions because ${\dot Q}^{(1)}(\tau)$ is a single-valued function of the instantaneous coupling $g(\tau)$, 
\begin{equation}
Q^{(1)}_j=    \frac{\hbar}{\Gamma}  \int_0^{2\pi} d\tau\, \dot{g}(\tau)\mathcal{Q}(\tau)  = \hbar \oint  dg \mathcal{Q}(g)= 0.
\end{equation}
Therefore the efficiency is basically  given by the ratio between the net-work produced and the zero-order heat current flowing into the system from the second thermal reservoir, namely, $\eta=W_{\rm cyc}^{\rm ext}/Q_2^{(0)}$. 
%We plot in \figref{fig: efficiency} the thermal efficiency as function of the losses ratio $\gamma_1/\gamma_2$ for three value of the adiabatic parameter $\zeta$ which, basically, is related to the coupling-modulation velocity. The in the plot clearly one sees that the efficiency increase when the loss asymmetry increases, as well as making faster the coupling-modulation. 

%\newpage 

\end{document}